\documentclass{jfm}
\usepackage{graphicx, color}
\usepackage{epstopdf, epsfig}
\usepackage{amsmath,amssymb,textcomp}

\usepackage{floatrow}
\usepackage{subfig}
\usepackage{multirow}

\usepackage{cleveref}
\crefname{section}{section}{sections}
\crefname{appsec}{appendix}{appendices}
\crefname{subsection}{subsection}{subsections}
\crefname{figure}{figure}{figures}
\crefname{table}{table}{tables}
\crefname{equation}{}{}
\Crefname{section}{Section}{Sections}
\Crefname{appsec}{Appendix}{Appendices}
\Crefname{subsection}{Subsection}{Subsections}
\Crefname{figure}{Figure}{Figures}
\Crefname{table}{Table}{Tables}

\definecolor{mygreen}{rgb}{0,.5,0}
\definecolor{mygray}{rgb}{.5,.5,.5}
\definecolor{mybrown}{rgb}{.43,.21,.1}

\usepackage{longtable}
\newcommand{\beq}{\begin{equation}}
\newcommand{\eeq}{\end{equation}}
\newcommand{\ov}{\overline}

\newcommand{\eps}{\varepsilon}
\newcommand{\blue}[1]{\textcolor{blue}{#1}}



\newcommand\solidrule[1][15pt]{\rule[0.5ex]{#1}{1pt}}
\newcommand\dashedrule{\mbox{%
	\solidrule[3pt]\hspace{3pt}\solidrule[3pt]\hspace{3pt}\solidrule[3pt]}}

\newcommand\dashdottedrule{\mbox{%
	\solidrule[4.5pt]\hspace{2pt}\solidrule[2pt]\hspace{2pt}\solidrule[4.5pt]}}

\shorttitle{High-wavenumber steady Rayleigh--B\'enard convection}
\shortauthor{B.\ Wen, A.\ Takla, D.\ Goluskin and G.\ P.\ Chini}

\title{Steady Rayleigh--B\'enard convection: strongly nonlinear high-wavenumber rolls}
\author{
\mbox{Baole Wen}\aff{1}\corresp{\email{bwen@nyit.edu, goluskin@uvic.ca}},
\mbox{Alexander Takla}\aff{2},
\mbox{David Goluskin}\aff{3}$\dagger$,
 \and
\mbox{Gregory P. Chini}\aff{4}}

\affiliation{
\aff{1}Department of Mathematics, New York Institute of Technology, Old Westbury, NY 11568, USA
\aff{2}Department of Physics, University of Michigan, Ann Arbor, MI 48109, USA
\aff{3}Department of Mathematics \& Statistics, University of Victoria, Victoria, BC, V8P 5C2, Canada
\aff{4}Program in Integrated Applied Mathematics, Department of Mechanical Engineering, University of New Hampshire, Durham, NH 03824, USA}

\begin{document}

\maketitle

\begin{abstract}

In Rayleigh--B\'{e}nard convection, two-dimensional steady rolls bifurcate supercritically at a Rayleigh number $Ra$ that depends on their horizontal-to-vertical aspect ratio $\Gamma$, and they exist at all larger $Ra$ despite being unstable. Heat transport by certain rolls---quantified by the Nusselt number $Nu$---closely resembles turbulent transport, yet $Nu$ scalings of rolls are understood only for specific boundary conditions and $\Gamma$--$Ra$ limits. Here we investigate the high-wavenumber limit $\Gamma = O(Ra^{-1/4})$ as $Ra \to \infty$, using numerics and matched asymptotic analysis. We compute steady rolls between stress-free boundaries for Prandtl numbers $10^{-1} \leq Pr \leq 10^{3/2}$ and $Ra$ reaching $10^{19}$. While the $\Gamma = O(Ra^{-1/4})$ limit gives smaller $Nu$ than when $\Gamma = O(1)$, we identify prefactors $c$ in $\Gamma = c\,Ra^{-1/4}$ that locally maximize $Nu$. These locally $Nu$-maximizing rolls display approximate scalings $Nu \propto Ra^{0.29}$ and $Re \propto Ra^{0.40}$, with the Reynolds number $Re$ defined using root-mean-square velocity. Our asymptotic analysis reveals a vertically stacked four-layer structure near each boundary, predicting $Nu = O(Ra^{3/10})$ and $Re = O(Ra^{2/5})$. This asymptotic construction largely follows that of Taylor vortices by \cite{Deguchi2023}, but we identify a thin plume region within the middle boundary layer whose inclusion eliminates the logarithmic factors in Deguchi's predictions. Asymptotic arguments and numerics suggest the same scalings for stress-free or no-slip boundaries, unlike in other $\Gamma$--$Ra$ limits. Our asymptotics extend the weakly nonlinear analysis of Blennerhassett \& Bassom (1994) into the strongly nonlinear regime and complement the asymptotics of Chini \& Cox (2009) for $\Gamma = O(1)$ rolls.

\end{abstract}

\vspace{-0.6in}

\section{Introduction}

Since its introduction by \citet{Rayleigh1916}, the canonical model for studying buoyancy-driven flow has been Rayleigh--B\'enard convection (RBC), where motion in a horizontal fluid layer is sustained by fixing a higher temperature on the bottom boundary than on the top one. The control parameters can be reduced to the dimensionless Rayleigh number $Ra$, which quantifies the effect of the imposed temperature difference across the layer; the Prandtl number $Pr$, which captures the relative dissipation rates of velocity gradients and temperature gradients; and the geometry of the domain. As $Ra$ is raised beyond the threshold for the onset of convection, flows become progressively more complex and eventually turbulent.

A central question about RBC is how mean quantities depend on $Ra$ and $Pr$, especially the Nusselt number $Nu$, which is the factor by which fluid flow amplifies heat transport across the layer relative to conduction. There is particular motivation to understand the asymptotic scaling of $Nu$ with $Pr$ and $Ra$ in the $Ra\to\infty$ limit because extreme $Ra$ values are typical in geophysical and astrophysical convection. For instance, $Ra$ values on the order of $10^{25}$ have been estimated for deep oceanic convection \citep{Chilla2012}, and astrophysical values can be far larger. The values of $Ra$ needed to unambiguously attain asymptotic scaling behaviours are seemingly beyond the reach of any direct numerical simulations or laboratory experiments to date \citep{Kadanoff2001,Ahlers2009,Chilla2012,Doering2020}.  
In light of this difficulty, a series of recent studies has aimed to determine large-$Ra$ asymptotic scalings of $Nu$ not for turbulent convection but instead for simpler steady flows that also solve the equations of motion.  Although steady flows are dynamically unstable at large $Ra$, they have been found to capture aspects of turbulent convection \citep{Waleffe2015,Sondak2015,Wen2015JFM,WenChini2018JFM,Wen2020JFM,Kooloth2021,Ding2021JFM,Wen2022JFM,Motoki2021,Motoki2022,Ouyang2025,Feng2025,He2026}. The scaling of $Nu$ with $Ra$ for certain steady flows is very similar to the scaling for turbulent convection, but the maximum $Nu$ among steady flows is larger than the corresponding turbulent values at all parameters where a direct comparison has been made \citep{Wen2020JFM,Wen2022JFM,Ding2021JFM,Motoki2022,He2026}. The relationship between steady and turbulent transport remains to be fully understood. A prerequisite to this understanding, which motivates our study, is to know how $Nu$ scales in asymptotic parameter limits for at least certain types of steady flows.

The present work concerns the simplest type of steady flows: two-dimensional (2D) convection rolls that bifurcate directly from the conduction state.  Considering a domain that is horizontally periodic or infinite, roll states consist of horizontally periodic counter-rotating pairs.  Such states can be uniquely labeled by the fundamental horizontal period $\Gamma$ (in units of layer height) and the number of rolls $m$ stacked vertically across the layer.  For each $(\Gamma,m)$, linear stability analysis of the conduction state gives a critical Rayleigh number $Ra_c$ \citep{Chandrasekhar1981}, independent of $Pr$, at which a branch of steady rolls bifurcates supercritically. While there are no exact formulas for the roll states at general $(Pr, Ra)$, computations indicate that each solution branch persists for all $Ra>Ra_c$ at every $Pr$, with $Nu$ increasing monotonically as $Ra$ is raised.  We confine our investigation to roll pairs that span the layer height (i.e., $m=1$).  Even for single pairs of layer-spanning rolls, however, the parameter-dependence $Nu(\Gamma,Pr,Ra)$ is far from fully understood.

Computations of steady rolls at large $Ra$ have been reported for top and bottom boundary conditions that are stress-free \citep{ChiniCox2009, Wen2020JFM} or no-slip \citep{Waleffe2015,Sondak2015,Wen2022JFM}, and there are both close parallels and key differences between the two cases. In the stress-free case, computations \citep{Wen2020JFM} as well as asymptotics \citep{ChiniCox2009} indicate $Nu\sim c_n(\Gamma) Ra^{1/3}$ as $Ra\to\infty$ with $\Gamma$ fixed, where the prefactor $c_n(\Gamma)$ is independent of $Pr$ in this limit and is maximized by $\Gamma\approx1.9$. In the no-slip case, computations \citep{Waleffe2015,Sondak2015,Wen2022JFM} indicate a different scaling: $Nu\sim c_n(\Gamma,Pr)Ra^{1/4}$ as $Ra\to\infty$ with $\Gamma$ and $Pr$ fixed. However, one can also consider simultaneous limits in which $\Gamma$ approaches zero or infinity at any rate between $\Gamma_{min}=O(Ra^{-1/4})$ and $\Gamma_{max}=O(Ra^{1/2})$ as $Ra\to\infty$, where $(\Gamma_{min},\Gamma_{max})$ is the interval of $\Gamma$ for which roll states exist, meaning that $Ra>Ra_c(\Gamma)$. In the no-slip case with $Pr=1$ and sufficiently large $Ra$, the dependence of $Nu$ on $\Gamma$ has a global maximum and one other local maximum, occurring at periods denoted by $\Gamma^*$ and $\Gamma^*_{loc}$, respectively. Rolls at each maximum become narrower (i.e., larger-wavenumber) as $Ra$ is raised, with numerics suggesting $\Gamma^*=O(Ra^{-1/5})$ and $\Gamma^*_{loc}=O(Ra^{-1/4})$, and that the global maximizers $\Gamma^*$ give $Nu=O(Ra^{1/3})$ \citep{Wen2022JFM}. In the stress-free case, steady rolls have not been computed previously outside of $\Gamma=O(1)$, and the scaling of $Nu$ in simultaneous limits of $\Gamma$ and $Ra$ has not been studied.

\begin{figure}[t]
\centering 
\includegraphics[width=0.98\textwidth]{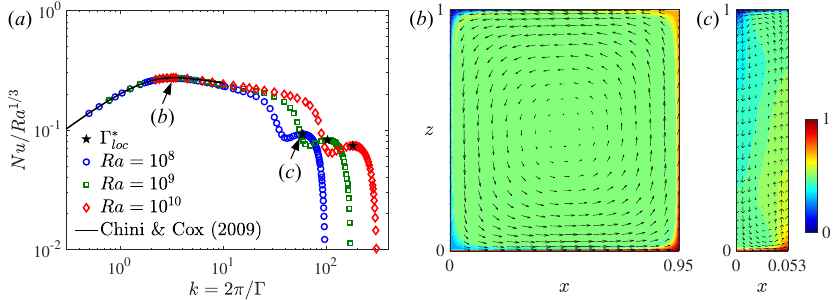}
\caption{(\emph{a}) Dependence of $Nu$, compensated by $Ra^{1/3}$, on the fundamental horizontal wavenumber $k=2\pi/\Gamma$ for steady convection rolls at $Pr=1$ between stress-free boundaries.  Lower-wavenumber solutions with $k \le 10$ are from \citet{Wen2020JFM}, and they agree quantitatively with the semi-analytical asymptotic construction of \citet{ChiniCox2009}. Stars indicate the secondary local maxima at $k=2\pi/\Gamma^*_{loc}$. (\emph{b}, \emph{c}) Rolls at $Ra=10^8$ with the globally $Nu$-maximizing period $\Gamma^*\approx1.9$ and the locally $Nu$-maximizing period $\Gamma^*_{loc}\approx 0.107$, respectively, that are marked in panel (\emph{a}). Colors represent the dimensionless temperature, ranging from 0 (cold) to 1 (hot), and arrows indicate the velocity vector field.  In panel (\emph{c}), the horizontal axis is scaled larger than the vertical axis for clarity.
}
\label{fig:Nu_loc}
\end{figure}

Here, we report computations of steady roll pairs in the stress-free case up to asymptotically large $Ra$, for various $10^{-1}\le Pr \le 10^{3/2}$ and over a period range $\Gamma_{min}\le\Gamma\le2\pi/10$ that broadens as $Ra\to\infty$ because $\Gamma_{min}=O(Ra^{-1/4})$. \Cref{fig:Nu_loc}(\emph{a}) shows the dependence of $Nu$ on $\Gamma$ at three different $Ra$ values, which is qualitatively representative of curves we have found at all $Pr$ when $Ra$ is sufficiently larger than $Ra_c$. As in the no-slip case \citep{Sondak2015,Wen2022JFM}, each curve in \cref{fig:Nu_loc}(\emph{a}) has a global maximum at a larger aspect ratio $\Gamma^*$ and a secondary local maximum at a smaller aspect ratio, which we find scales as $\Gamma^*_{loc}=O(Ra^{-1/4})$. Panels (\emph{b}) and (\emph{c}) show examples of the global and local maximizers, respectively. The focus of the present work is rolls whose aspect ratios scale as $O(Ra^{-1/4})$, including with the exact $\Gamma^*_{loc}$ values. We therefore compute rolls at $\Gamma$ values near $\Gamma^*_{loc}$ over large ranges of $Ra$, including parameter sweeps where $\Gamma = c\,Ra^{-1/4}$ with constant $c$.

Complementing our numerics, we carry out a matched asymptotic analysis in which the domain is decomposed into four vertically stacked layers, one of which is divided horizontally into two regions. The flow in each of the five regions is described by a different asymptotic reduction of the governing partial differential equations (PDEs). Numerical solution of the reduced PDEs is left for future work, but scalings can be anticipated from the matching conditions between solutions in adjacent regions. The resulting predictions for $Nu$ and for the Reynolds number $Re$, defined using root-mean-square velocity, are that $Nu \sim c_n\,Ra^{3/10}$ and $Re\sim c_r\,Ra^{2/5}$ as $Ra\to\infty$ with $\Gamma = c\,Ra^{-1/4}$, where the prefactors $c_n$ and $c_r$ generally depend on $c$ and $Pr$.

Our asymptotic construction largely follows the analysis of steady axisymmetric Taylor vortices in Taylor--Couette flow by \citet{Deguchi2023}, who examined the limit $\Gamma = c\,Ra^{-1/4}$ along with several other $\Gamma$--$Ra$ limits. Axisymmetric Taylor--Couette flow with an asymptotically narrow gap between cylinders is mathematically equivalent to 2D RBC at $Pr=1$ \citep{Eckhardt2020}. Here, we argue that RBC rolls satisfying  $\Gamma = c\,Ra^{-1/4}$ exhibit identical transport scalings under both stress-free and no-slip boundary conditions. However, we propose that \citeauthor{Deguchi2023}'s solution structure for the $\Gamma = c\,Ra^{-1/4}$ limit is incomplete, as certain matching requirements cannot be satisfied. To address this, we introduce another region into the asymptotic structure---an addition that is also supported by our numerical results. The resulting predictions differ slightly from those of \citeauthor{Deguchi2023} in that our scalings do not have logarithmic corrections. Our predictions for the different regions are supported by numerical computations with $Ra$ up to $10^{19}$. Like the asymptotic construction for $\Gamma=O(1)$ rolls by \citet{ChiniCox2009}, our analysis is strongly nonlinear.  It extends the weakly nonlinear analysis of \citet{Blennerhassett1994}, who also considered the $\Gamma= O(Ra^{-1/4})$ regime but only for $\Gamma$ asymptotically close to $\Gamma_{min}$. 

The rest of this paper is organized as follows. \Cref{sec:formulation} gives the governing PDEs and describes the numerical methods used to compute steady solutions regardless of dynamical stability. \Cref{sec:LocaMax_numerics} presents rolls of locally $Nu$-maximizing period $\Gamma^*_{loc}$ that we have computed up to $Ra = 10^{16}$. \Cref{sec:LocaMax_analysis} develops a matched asymptotic analysis of such rolls and presents additional computations that support the asymptotic predictions with $Ra$ as large as $10^{19}$. \Cref{sec:fixed Gamma} illustrates how rolls transition between different asymptotic regimes as we sweep through $Ra$ at small but fixed $\Gamma$ values, including the change in flow structure from the solutions of \citet{Blennerhassett1994} to strongly nonlinear high-wavenumber rolls. \Cref{sec:conclusions} offers conclusions, \cref{app:BB} details the relation between our asymptotics and those of \citet{Blennerhassett1994}, and \cref{appB:Data} gives numerical properties for many of the rolls we have computed.

\section{Governing equations and numerical methods}
\label{sec:formulation}

We consider a 2D fluid layer in a domain $(x, z)\in[0,\Gamma]\times[0,1]$ that has been nondimensionalized so that the vertical ($z$) extent is unity and the horizontal ($x$) direction is $\Gamma$-periodic. In dimensionless form, the Boussinesq equations generally used to model RBC can be expressed as
\begin{subequations}
\label{eq: bouss}
\begin{align}
{\partial_t \mathbf{u}} + \mathbf{u}\cdot\nabla{\mathbf{u}} &= -\nabla{p} + {(Pr/Ra)^{1/2}} \nabla^2\mathbf{u} + T\mathbf{\hat{z}}, \label{Momentum}\\
 \nabla\cdot{\mathbf{u}} &= 0,\label{Continuity}\\
{\partial_t T} + \mathbf{u}\cdot\nabla T &= {(Pr Ra)^{-1/2}}\nabla^2T, \label{Energy}
\end{align}
\end{subequations}
where $\mathbf{u} = u\mathbf{\hat{x}} + w\mathbf{\hat{z}}$ is the velocity field, $p$ is the pressure and $T$ is the temperature. Lengths have been scaled by the dimensional layer height $h$, temperature has been scaled by the temperature drop $\Delta$ from the bottom boundary to the top one, and time ($t$) has been scaled by $h/U_f$, where $U_f = \sqrt{g\alpha h \Delta}$ is the free-fall velocity for the given thermal expansion coefficient $\alpha$ and gravitational acceleration $g$ acting in the $-\mathbf{\hat z}$ direction. Along with the aspect ratio $\Gamma$, the dimensionless parameters are the Prandtl number $Pr=\nu/\kappa$ and the Rayleigh number $Ra={g\alpha h^3\Delta}/{(\kappa\nu)}$, where $\nu$ is the kinematic viscosity and $\kappa$ is the thermal diffusivity.

At the top and bottom boundaries we impose the same isothermal and stress-free conditions as \citet{Rayleigh1916}, which in dimensionless form are 
\begin{equation}
\label{eq:BCs}
T|_{z=0}=1,\quad T|_{z=1}=0,\quad w|_{z=0,1}=0,\quad \partial_z u|_{z=0,1}=0.
\end{equation}
Branches of steady roll pairs with fundamental period $\Gamma$ bifurcate supercritically from the conduction state as $Ra$ increases past a $Pr$-independent critical value $Ra_c$. For the given boundary conditions, \citeauthor{Rayleigh1916}'s linear stability analysis of the conduction state gives \citep{Chandrasekhar1981}
\begin{equation}
Ra_c = \frac{(\pi^2+k^2)^3}{k^2},
\label{eq:Ra_c}
\end{equation}
where $k=2\pi/\Gamma$ is the fundamental horizontal wavenumber. The minimum $Ra_c$ value of $4\pi^4/27$ is attained by rolls with $\Gamma=2\sqrt2$, and at all $Ra>4\pi^4/27$ there exists a roll state for every $\Gamma\in(\Gamma_{min},\Gamma_{max})$, where $\Gamma_{min}\sim2\pi Ra^{-1/4}$ and $\Gamma_{max}\sim 2\pi^{-2}Ra^{1/2}$, as $Ra\to\infty$.

On a 2D domain one can express~\cref{eq: bouss} in terms of the stream function $\psi$, defined by $\mathbf{u} = \partial_z \psi \mathbf{\hat x} - \partial_x \psi \mathbf{\hat z}$, and the (negative) scalar vorticity $\omega=\partial_x w- \partial_z u = -\nabla^2 \psi$. Steady states ($\partial_t=0$) in the vorticity--stream function formulation obey
\begin{subequations}
\label{eq: steadybouss}
\begin{align}
\partial_z\psi\partial_x\omega - \partial_x\psi\partial_z\omega  &= 
({Pr}/{Ra})^{1/2}\; \nabla^2 \omega + \partial_x T, \label{eq: omega} \\
\nabla^2 \psi &= -\omega, \label{eq: psi}\\
\partial_z\psi\partial_xT - \partial_x\psi\partial_zT  & = (Pr Ra)^{-1/2}\; \nabla^2 T. \label{eq: T}
\end{align}
\end{subequations}
The zero normal-flow and stress-free conditions on $w$ and $u$ in~\cref{eq:BCs} are enforced by imposing $\psi=0$ and $\omega=0$, respectively, along the top and bottom boundaries. Although~\cref{eq:BCs} alone does not require $\psi$ to have the same value at the top as at the bottom, fixing $\psi=0$ at both boundaries amounts to choosing the reference frame with no mean horizontal velocity.

Numerical solution of~\cref{eq: steadybouss} is carried out using the code developed for \citet{Wen2020JFM}. The code implements a slightly different form of~\cref{eq: steadybouss} in which the temperature field is represented not by $T$ but by the deviation of $T$ from the conduction profile $1-z$.  These PDEs appear as (2.4) in \citet{Wen2020JFM}, for instance, wherein the deviation $T-(1-z)$ is denoted by $\theta$. (In the present paper we reserve $\theta$ for different decompositions of the $T$ field.) The code was modified slightly to employ the nondimensionalization used in~\cref{eq: steadybouss} but is otherwise unchanged. Details of our numerical algorithm can be found in \cite{Wen2020JFM}. Briefly, the PDEs are discretized using a Fourier series in $x$ and a Chebyshev collocation method in $z$ \citep{Boyd2000, Trefethen2000}. The discretized equations are a quadratic algebraic system, which is solved iteratively using a Newton--GMRES (generalized minimal residual) scheme.

The spatial averages that we analyze are the Nusselt and Reynolds numbers. The Nusselt number $Nu$ is defined as the ratio of total vertical heat flux to the flux carried by conduction alone. The Reynolds number is $Re=hU_{rms}/\nu$, where $U_{rms}$ is the dimensional root-mean-square velocity of the flow. In terms of the dimensionless variables in~\cref{eq: bouss} and~\cref{eq: steadybouss},
\begin{subequations}
\label{eq: Nu Re}
\begin{align}
Nu &= 1 + (PrRa)^{1/2}\langle wT\rangle=1- (PrRa)^{1/2}\langle T\partial_x \psi\rangle, \label{eq: Nu def} \\
Re &= (Ra/Pr)^{1/2}\langle |\mathbf{u}|^2\rangle^{1/2} = (Ra/Pr)^{1/2}\langle |\nabla\psi|^2\rangle^{1/2},\label{eq: Re def} 
\end{align}
\end{subequations}
where $\langle\cdot\rangle$ denotes a spatial average. Our asymptotics make use of an equivalent expression for $Nu$ as a horizontal average,
\begin{equation}
\label{eq:Nu(z) small x}
Nu =- \partial_z \overline{T} + (PrRa)^{1/2} \overline{wT}  = - \partial_z \overline{T} - (PrRa)^{1/2} \overline{T\partial_x\psi}
\end{equation}
for any $z\in[0,1]$, where an overline denotes a horizontal average. The fact that this expression is $z$-independent is shown by integrating~\cref{Energy} or~\cref{eq: T} in $x$ and then integrating by parts to show that the $z$-derivative of~\cref{eq:Nu(z) small x} is zero. A further  integration of~\cref{eq:Nu(z) small x} in $z$ confirms its equivalence to the volume-averaged expression~\cref{eq: Nu def} for $Nu$.

At fixed $Ra$ and $Pr$, we compute steady rolls at various $\Gamma$ and use cubic spline interpolation to determine precisely the $O(Ra^{-1/4})$ period $\Gamma^*_{loc}$ that locally maximizes $Nu$. \Cref{fig:Nu_loc} illustrates that this local maximum is clearly distinguished from the global maximum at $\Gamma\approx1.9$ when $Ra$ is sufficiently large. At most $Pr$ and $Ra$ we do not complete a full sweep over $(\Gamma_{min},\Gamma_{max})$ like the ones shown in \cref{fig:Nu_loc}; instead we choose $\Gamma$ values only as needed to find $\Gamma^*_{loc}$.

\section{Rolls of locally $Nu$-maximizing period: Numerical solutions}
\label{sec:LocaMax_numerics}

We have computed steady rolls with the locally $Nu$-maximizing horizontal period $\Gamma^*_{loc}$ for Rayleigh numbers over the range $10^{8} \le Ra \le 10^{16}$ and Prandtl numbers over the range $10^{-1} \leq Pr \leq 10^{3/2}$. When $Pr\lesssim 10^{-5/4}$ we find that the secondary local maximum of $Nu$ does not exist, at least up to $Ra=10^{14}$, meaning that curves like those in \cref{fig:Nu_loc} have only the maximum near $\Gamma\approx1.9$, so $\Gamma^*_{loc}$ is undefined.

\subsection{Spatial averages}
\label{sec:spatial_averages}

\begin{figure}[t!]
\centering
\includegraphics[width=0.98\textwidth]{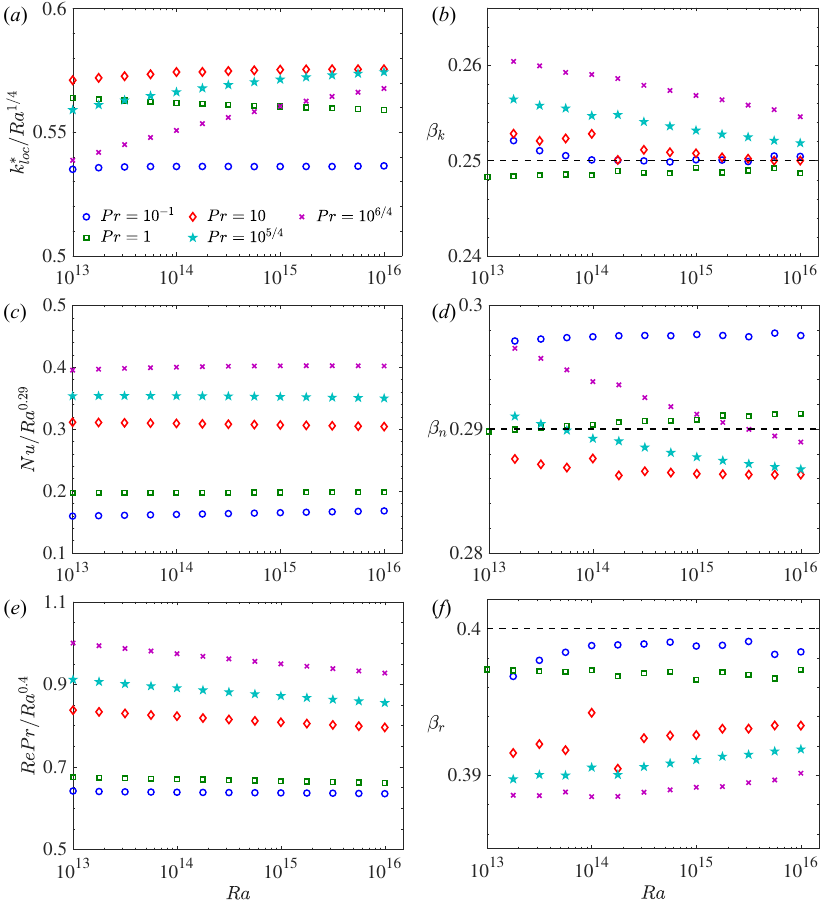}
\caption{Mean properties of rolls with locally $Nu$-maximizing horizontal periods $\Gamma^*_{loc}$ for a range of $Ra$ at five $Pr$ values. The top row shows ($a$)~these periods in terms of their fundamental horizontal wavenumber $k^*_{loc} = 2\pi/\Gamma^*_{loc}$, compensated by the $Ra^{1/4}$ scaling that is the limit of $k$ studied asymptotically in \S~\ref{sec:LocaMax_analysis}, and ($b$)~best-fit local scaling exponents for $k^*_{loc}\approx c_k Ra^{\beta_k}$, which are estimated as finite-difference approximations of $\mathrm{d}(\log k^*_{loc})/\mathrm{d}(\log Ra)$. The middle row shows ($c$)~the locally maximal $Nu$ values, compensated by the numerically approximated scaling $Ra^{0.29}$, and ($d$)~best-fit local scaling exponents for $Nu\approx c_n Ra^{\beta_n}$. For these same rolls the last row shows ($e$)~the $Re$ values, compensated by the numerically approximated scaling $Ra^{0.40}$, and ($f$)~best-fit local scaling exponents for $Re\approx c_r Ra^{\beta_r}$. Dashed lines in the right-hand plots indicate the scaling exponents by which the corresponding left-hand plots are compensated. The plotted data are tabulated in \cref{appB:Data} along with values for $Pr = 10^{1/4}$, $10^{1/2}$ and $10^{3/4}$ that are not plotted for clarity.}
\label{fig:NuLocPrfixed}
\end{figure}

For the rolls we computed with periods of $\Gamma^*_{loc}$ at various fixed $Pr$, \cref{fig:NuLocPrfixed} shows the $Ra$-dependence of the periods, plotted in terms of the fundamental wavenumber $k^*_{loc}=2\pi/\Gamma^*_{loc}$, along with the $Ra$-dependence of $Nu$ and $Re$ for these rolls. In panel ($a$) the $k^*_{loc}$ values are compensated by $Ra^{1/4}$, so the approach to horizontal lines suggests $k^*_{loc}\sim c_k(Pr)Ra^{1/4}$ asymptotically. In other words, rolls with the locally $Nu$-maximizing period $\Gamma^*_{loc}$ narrow at the same $O(Ra^{-1/4})$ rate as the minimum period $\Gamma_{min}$. Measuring the local scaling exponents between data points in panel ($a$) gives even clearer evidence that the exponents approach $1/4$; panel ($b$) shows local exponents for the same data as in panel ($a$). Rolls at larger $Pr$ values are farther from exhibiting a clean asymptotic scaling because they apparently require larger $Ra$ to reach this regime, as also found when $\Gamma=O(1)$ instead of $\Gamma=O(Ra^{-1/4})$ \citep{Wen2020JFM}. Extrapolating the data series in panel ($a$) to horizontal asymptotes suggests that the prefactors $c_k(Pr)$ increase monotonically by about 10\% as $Pr$ is raised from $10^{-1}$ to $10^{3/2}$.

For the rolls with periods $\Gamma^*_{loc}$, the $Ra$-dependence of $Nu$ and $Re$ is shown in the left-hand panels ($c$) and ($e$) of \cref{fig:NuLocPrfixed}, along with the local scaling exponents in right-hand panels ($d$) and ($f$), respectively. The right-hand plots suggest that, for all five $Pr$ values, the asymptotic exponents of $Nu\sim c_n\,Ra^{\beta_n}$ and $Re\sim c_r\,Ra^{\beta_r}$ lie in the ranges $\beta_n\in[0.28,0.30]$ and $\beta_r\in[0.39,0.40]$. Based on these measured exponents, the left-hand plots of $Nu$ and $Re$ have been compensated by $Ra^{0.29}$ and $Ra^{0.40}$, respectively, thus their data series are nearly horizontal. These measured exponents are close to the exact values suggested by the asymptotics in \cref{sec:LocaMax_analysis} below, but they are not identical. The asymptotics also suggest that the scaling exponents are eventually $Pr$-independent, as in the $\Gamma=O(1)$ case \citep{Wen2020JFM}, whereas our computations up to $Ra=10^{16}$ cannot rule out weak $Pr$-dependence in the exponents. The $Pr=0.1$ data, which seems to be closest to clean asymptotic behaviour, suggests exponents of $\beta_n\approx 0.298$ and $\beta_r\approx0.399$. Remarkably, all three digits of these values agree with the exact exponents predicted by our asymptotics below.

The rolls represented in \cref{fig:NuLocPrfixed} can be compared, at least when $Pr=1$, to analogous computations with no-slip boundary conditions \citep{Wen2022JFM}. This comparison suggests that $\Gamma^*_{loc}=O(Ra^{-1/4})$ whether boundaries are stress-free or no-slip, and that rolls of these horizontal periods have the same asymptotic scaling exponents of $Nu$ and $Re$. In both cases, the local scaling exponents to two digits are $\beta_n\approx0.29$ and $\beta_r\approx 0.40$ when $Pr=1$. The apparent agreement between the two boundary conditions when $\Gamma=O(Ra^{-1/4})$ is in contrast to other limits such as $\Gamma=O(1)$, where scaling exponents clearly differ between the stress-free and no-slip cases.

\subsection{Spatial structure}

\begin{figure}[t]
\centering
\includegraphics[width=0.84\textwidth]{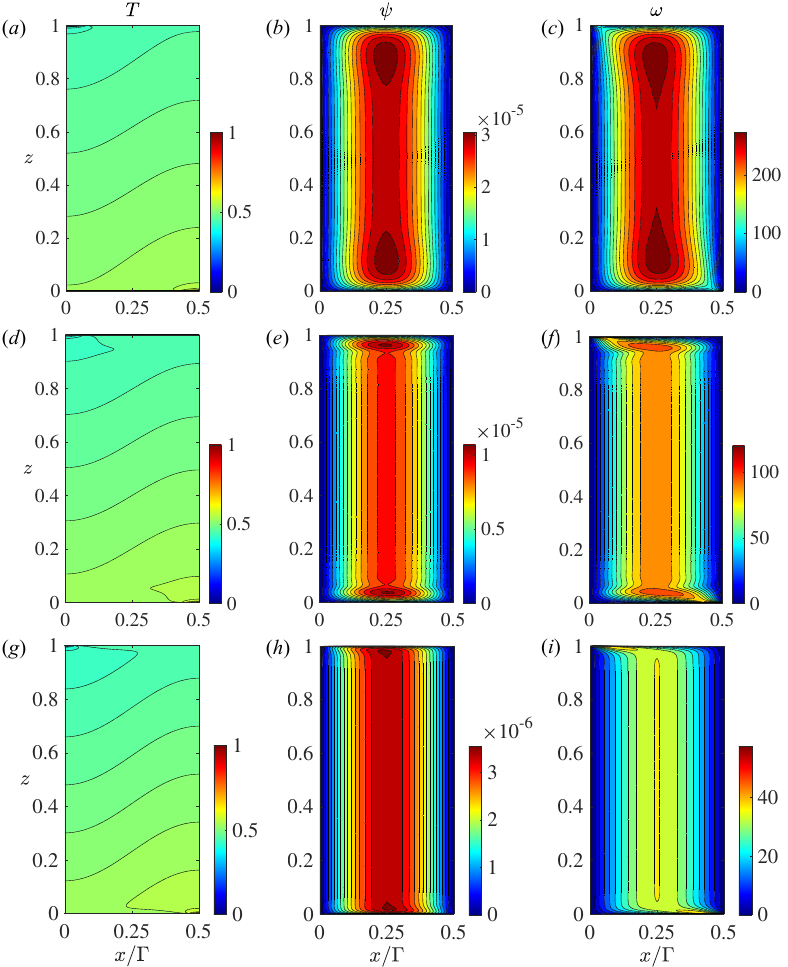}
\caption{Contours of $T$ (left), $\psi$ (middle) and $\omega$ (right) for steady rolls of locally $Nu$-maximizing period $\Gamma^*_{loc}$ at $Ra=10^{15}$ with $Pr=0.1$ (top), $Pr=1$ (middle) and $Pr=10$ (bottom). From top to bottom, $\Gamma^*_{loc}\approx0.00208$, 0.00199 and 0.00194, and the maximum $\psi$ values are $3.17\times10^{-5}$, $1.08\times10^{-5}$ and $3.59\times10^{-6}$ in the free-fall units of~\cref{eq: steadybouss}.}
\label{fig:Contour_TPsiOmega}
\end{figure}

We turn now to the spatial structure of rolls whose $Nu$-maximizing periods $\Gamma^*_{loc}$ have been reported in \cref{sec:spatial_averages}. \Cref{fig:Contour_TPsiOmega} shows the temperature, stream function and vorticity fields for steady rolls with $Ra=10^{15}$ and $Pr = 0.1$, 1 or 10. This $Ra$ value is close to the asymptotic regime, although least so when $Pr=10$, as can be seen from the $Nu$ and $Re$ scalings in \cref{fig:NuLocPrfixed}. At each $Pr$, at least three distinct spatial regions can be discerned. The core region, far from the top and bottom boundaries, has almost purely vertical velocity, which correspond to $\psi$ and $\omega$ being nearly $z$-independent. The thermal boundary layers at the top and bottom are hard to discern in \cref{fig:Contour_TPsiOmega} but are examined below. In between the core and the thermal boundary layers are near-wall recirculation regions, visible as closed streamlines in \cref{fig:Contour_TPsiOmega}. When $Pr$ is larger, the recirculation region is smaller, but pre-asymptotic effects are stronger also. The three regions identified in our numerically computed rolls are consistent with the asymptotic analysis of \cref{sec:LocaMax_analysis}, which refines the roll solution into still more regions.

The rolls of period $\Gamma^*_{loc}$ depicted in \cref{fig:Contour_TPsiOmega} are quite similar to 2D incompressible, wall-to-wall optimal velocity fields designed to maximize heat flux between the boundaries when the flow is only subject to an enstrophy budget rather than obeying the Boussinesq equations \citep{Hassanzadeh2014,Souza2020}. The designed flows are also horizontally periodic rolls with a heat-exchanger core (defined in \cref{Numerics_Core} below), thermal boundary layers, and recirculation regions in between.  While the design problem proposed by \citet{Hassanzadeh2014} has no $Ra$ since there is no buoyancy, for the stress-free boundary conditions, its candidate maximizers of heat flux become narrower with increasing enstrophy budget according to the aspect ratio scaling $\Gamma = O(Pe^{-6/17})$.  In their study, the P\'{e}clet number---defined via $Pe^2 = \langle |\nabla\mathbf{u}|^2\rangle$ with $\mathbf{u}$ non-dimensionalised using the thermal diffusion velocity scale---serves as the fixed-enstrophy constraint.  When applied to Rayleigh--B\'enard convection, where $Pe^2 = Ra(Nu - 1)$, \cite{Hassanzadeh2014} demonstrate  that this $Nu$-maximizing aspect ratio maps precisely to $\Gamma = O(Ra^{-1/4})$, matching the behaviour of our RBC rolls of period $\Gamma^*_{loc}$.  Yet, a key distinction remains: the optimized rolls designed by \citet{Hassanzadeh2014} corresponding to $\Gamma = O(Ra^{-1/4})$ achieve a maximal heat transport of $Nu = 1 + 0.1152Ra^{5/12}$, in agreement with the analytical upper bound scaling $Nu \lesssim Ra^{5/12}$ derived by \citet{Whitehead2011}.  In contrast, the RBC rolls of period $\Gamma^*_{loc} = O(Ra^{-1/4})$---which fully satisfy the Boussinesq equations---yield a lower heat flux scaling of $Nu \propto Ra^{0.29}$ based on our numerical results.

\subsubsection{Core region}\label{Numerics_Core}

In the core region, the rolls in \cref{fig:Contour_TPsiOmega} have a simple structure. In addition to $\psi$ and $\omega$ being nearly $z$-independent, these fields and $T$ have horizontal variation that is dominated by the lowest Fourier mode. This suggests that the core flow can be approximated by $\psi=\tilde{A}\sin {kx}$ for some constant $\tilde{A}$. The governing PDEs~\cref{eq: steadybouss}, but not the associated boundary conditions, are satisfied by this $\psi$ and its corresponding $\omega$ and $T$,
\begin{equation}
\label{eq:heatexchanger}
\psi = \tilde{A}\sin{kx}, \;\; \omega = \tilde{A}k^2\sin{kx}, \;\;T = \tfrac{1}{2} - k^4Ra^{-1}\left(z-\tfrac12\right)-\tilde{A}\left(\tfrac{Pr}{Ra}\right)^{1/2} k^3 \cos{kx}.
\end{equation}
The constant part $\tfrac12(1+k^4Ra^{-1})$ of the $T$ field in~\cref{eq:heatexchanger} does not affect the PDEs; it is chosen to give $T$ the desired symmetry when rotated 180$^\circ$ about the midpoint of the roll. In~\cref{eq:heatexchanger}, vertical advection of temperature in the up- and downwelling regions balances the horizontal diffusion of temperature between these regions. The same structure arises in the core region of porous medium RBC \citep{Hewitt2012, Wen2015JFM}, and it has been called a ``heat-exchanger flow''. To show more precisely that the core regions of the rolls in \cref{fig:Contour_TPsiOmega} are well described by~\cref{eq:heatexchanger}, we consider the horizontal decompositions
\begin{equation}
\label{eq:decomposition}
\begin{bmatrix} 
\omega(x,z) \\
\psi(x,z) \\
T(x,z)
\end{bmatrix}
= 
\begin{bmatrix} 
\overline{\omega}(z) \\
\overline{\psi}(z) \\
\overline{T}(z)
\end{bmatrix}
+ \sum_{n=1}^\infty
\begin{bmatrix} 
\hat{\omega}_n(z)\sin{(nkx)} \\
\hat{\psi}_n(z)\sin{(nkx)} \\
\hat{\theta}_n(z)\cos{(nkx)}
\end{bmatrix}.
\end{equation}
Here, overlines denote horizontal averages over a \emph{single} roll. Note that the mean temperature profile $\overline{T}(z)$ is the same for both rolls in a counter-rotating pair, and for the heat-exchanger flow~\cref{eq:heatexchanger} the mean stratification $\partial_z\overline{T}= - k^4Ra^{-1}$ independently of the unknown amplitude $\tilde{A}$. In contrast, the mean profiles $\overline{\psi}(z)$ and $\overline{\omega}(z)$ have opposite signs for each roll in a pair, and for a heat-exchanger flow they have constant values of $2\tilde{A}/\pi$ and $2k^2\tilde{A}/\pi$, respectively.

\begin{figure}[t]
\centering
\includegraphics[width=0.96\textwidth]{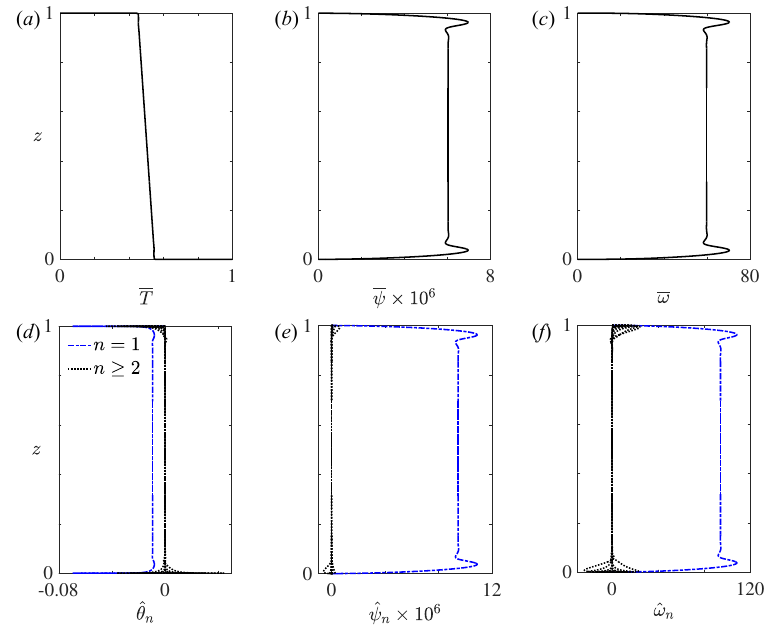}
\caption{Components of the horizontal Fourier decomposition~\cref{eq:decomposition} for the steady counterclockwise roll with $Ra=10^{15}$, $Pr = 1$ and $\Gamma = \Gamma^*_{loc}\approx 0.0019938$ that is shown in the middle row of \cref{fig:Contour_TPsiOmega}. The slope of $\overline{T}(z)$ in the interior agrees to five digits with the value $\partial_z\overline{T}=-k^4Ra^{-1}\approx-9.8626\times 10^{-2}$ of the heat-exchanger flow~\cref{eq:heatexchanger} at this $\Gamma$.}
\label{fig:Spectra_TPsiOmega}
\end{figure}

\Cref{fig:Spectra_TPsiOmega} shows all $z$-dependent profiles in the expansion~\cref{eq:decomposition} for the $Pr=1$ roll depicted in the middle row of \cref{fig:Contour_TPsiOmega}. The top row shows the horizontal averages $\ov{T}$, $\ov{\psi}$ and $\ov{\omega}$, while the bottom row shows the Fourier coefficients $\theta_n$, $\psi_n$ and $\omega_n$ for $1\le n\le 10$. (Analogous plots for the $Pr=0.1$ and $Pr=10$ rolls in \cref{fig:Contour_TPsiOmega} are similar.) The mean profiles in the top row of \cref{fig:Spectra_TPsiOmega} are consistent with the core region being a heat-exchanger flow~\cref{eq:heatexchanger}: the midline temperature gradient $\partial_z\overline{T}(1/2)\approx -0.0986258$ agrees well with the heat-exchanger value $-k^4Ra^{-1}\approx-0.0986264$, and the midline ratio $\ov{\omega}/\ov{\psi}(1/2)\approx9.93108\times10^6$ agrees exactly with $k^2\approx9.93108\times10^6$. Heat-exchanger structure is further confirmed by the bottom row of \cref{fig:Spectra_TPsiOmega}, which shows that $n=1$ Fourier modes of $T$, $\psi$ and $\omega$ dominate all $n\ge2$ modes in the core region. This heat-exchanger core persists at all $Pr$ and sufficiently large $Ra$ for which we have computed rolls of period $\Gamma^*_{loc}$, as well as at nearby $\Gamma$. The departure from heat-exchanger structure at other $\Gamma$ is analyzed in \S~\ref{sec:fixed Gamma}.

\subsubsection{Thermal boundary layers}

Thermal boundary layers are evident upon examining $\overline{T}(z)$ near the boundaries. The temperature change across these layers is $O(1)$ at large $Ra$ in the dimensionless units of~\cref{eq: steadybouss}, and both convection and conduction contribute significantly to the total heat flux.
These facts together imply that, as $Ra\to\infty$, the layer thickness decreases as $O(1/Nu)$, which is approximately as $Ra^{-0.29}$ according to the measured $Nu$ scalings reported in \S~\ref{sec:spatial_averages}. \Cref{fig:MeanBL_TPsiOmega}($a$) shows $\overline{T}$ plotted against the scaled coordinate $z/Ra^{-0.29}$ for three different $Ra$ values with $Pr=1$. The shape of the thermal boundary layer collapses to a single asymptotic curve at all three $Ra$. Similar collapse to slightly different curves occurs when $Pr=0.1$ and $10$.

\subsubsection{Near-wall recirculation regions}

\begin{figure}[t]
\centering
\includegraphics[width=0.96\textwidth]{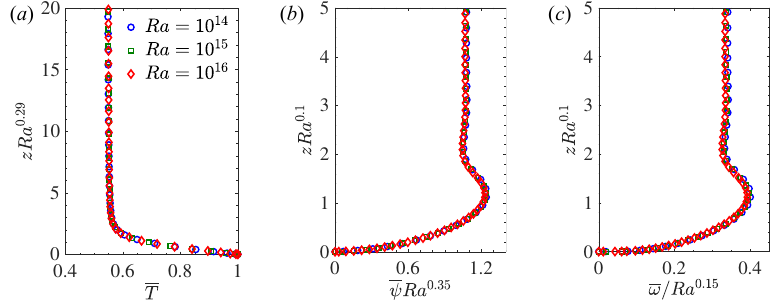}
\caption{Horizontally averaged profiles of ($a$) temperature, ($b$) stream function and ($c$) vorticity near the lower boundary, with the values and the $z$ coordinate rescaled so that the profiles for $Ra=10^{14}$, $10^{15}$ and $10^{16}$ with $Pr=1$ collapse onto single curves. The $Nu$-maximizing periods at these $Ra$ values are $\Gamma^*_{loc}\approx0.00354$, 0.00199 and 0.00112, respectively. For clarity, $\overline{\psi}(z)$ and $\overline{\omega}(z)$ values are plotted at only a subset of the collocation points.}
\label{fig:MeanBL_TPsiOmega}
\end{figure}

The near-wall recirculation regions can be identified not only by the closed streamlines in \cref{fig:Contour_TPsiOmega} but by coinciding boundary layers in $\overline{\psi}(z)$ and $\overline{\omega}(z)$. These velocity boundary layers are thicker than the thermal ones, and their thickness decreases relatively slowly as $Ra\to\infty$. The measured thickness of the velocity boundary layers---i.e., of the recirculation regions---scales approximately as $Ra^{-0.10}$, in contrast with approximately $Ra^{-0.29}$ for the thermal boundary layer. The magnitudes of $\psi$ and $\omega$ in the recirculation regions scale like their core values. Evaluating $\overline{\psi}(1/2)$ directly gives the $\psi$ scaling as approximately $Ra^{-0.35}$, and multiplying by $k^2$ according to \cref{eq:heatexchanger} gives the $\omega$ scaling as approximately $Ra^{0.15}$. \Cref{fig:MeanBL_TPsiOmega}($b$, $c$) show $\overline{\psi} Ra^{0.35}$ and $\overline{\omega}/Ra^{0.15}$ plotted against the scaled coordinate $z/Ra^{-0.10}$ for three different $Ra$ values with $Pr=1$. These velocity boundary layers collapse to asymptotic curves for $\overline{\psi}$ and for $\overline{\omega}$ at all three $Ra$, and analogous collapse to corresponding curves occurs at $Pr=0.1$ and 10.

\section{Rolls of locally $Nu$-maximizing period: Asymptotic analysis}
\label{sec:LocaMax_analysis}

The numerical computations of steady rolls presented in \cref{sec:LocaMax_numerics} suggest that the locally $Nu$-maximizing period scales as $\Gamma^*_{loc}=O(Ra^{-1/4})$. In the present section, we perform a matched asymptotic analysis of rolls in limits where $\Gamma=c\,Ra^{-1/4}$ as $Ra\to\infty$ for fixed $Pr$, which includes the $c$ value giving $\Gamma^*_{loc}$ as well as other $c$ values. Equivalently, the fundamental wavenumber of the horizontal period scales as $k=O(Ra^{1/4})$. Our analysis is guided by the numerical results of \cref{sec:LocaMax_numerics} and by the asymptotic analysis of steady axisymmetric Taylor--Couette flow by \citet{Deguchi2023}, which we argue must be modified by an additional region to allow for matching of solutions between all regions. A main takeaway of our analysis is an exact prediction for the exponents $\beta_n$ and $\beta_r$ in the asymptotic scalings $Nu\sim c_n\,Ra^{\beta_n}$ and $Re \sim c_r\,Ra^{\beta_r}$ for rolls with periods $\Gamma^*_{loc}$. Fitting to the numerical data from \cref{sec:LocaMax_numerics} suggests $0.28\lesssim \beta_n\lesssim 0.30$ and $0.39\lesssim\beta_r\lesssim0.40$, depending on $Pr$, but even at $Ra=10^{16}$ this data has not reached clean asymptotic scaling.  Our analysis suggests that the exact asymptotic exponents are $\beta_n=3/10$ and $\beta_r=2/5$.

To study the limit in which rolls get narrower as $\Gamma=O(Ra^{-1/4})$, we set $\Gamma = Ra^{-1/4}L$ with $L$ constant. We let $x = Ra^{-1/4} X$, so the scaled horizontal domain is $X\in[0, L]$, with corresponding fundamental horizontal wavenumber $K = 2\pi/L = Ra^{-1/4}k$. In the asymptotics it is convenient to decompose $T$ as
\beq
T(X,z) = \tau(z) + \theta(X,z),
\label{eq:T decomp}
\eeq
where $\tau(z)$ is constant or linear in $z$. In the core, $\tau$ is chosen to agree with the horizontal average $\ov{T}$ to leading order, while in the boundary layers, $\tau$ is chosen to be a constant value that does not need to coincide with $\ov{T}$.  In $(X,z)$ coordinates, and with the $T$ decomposition~\cref{eq:T decomp} for linear $\tau$, the governing PDEs~\cref{eq: steadybouss} are
\begin{subequations}
\label{eq: steadyboussX}
\begin{align}
\partial_z\psi\partial_X\omega - \partial_X\psi\partial_z\omega  &= 
Pr^{1/2}{Ra}^{-3/4}\left(Ra^{1/2}\partial_X^2 + \partial_z^2\right) \omega + \partial_X\theta, \label{eq: omegaX} \\
\left(Ra^{1/2}\partial_X^2 + \partial_z^2\right) \psi &= -\omega, \label{eq: psiX}\\
\partial_z\psi\partial_X\theta - \partial_X\psi\partial_z(\tau+\theta) & =   Pr^{-1/2}{Ra}^{-3/4} \left(Ra^{1/2}\partial_X^2 + \partial_z^2\right)\theta. \label{eq: thetaX}
\end{align}
\end{subequations}
In $(X,z)$ coordinates, the expression~\cref{eq:Nu(z) small x} for $Nu$ as a horizontal average is
\begin{equation}
\label{eq:Nu(z)}
Nu = - \partial_z \overline{T} - Pr^{1/2}Ra^{3/4} \overline{T\partial_X\psi} 
\end{equation}
at each $z$, where in these variables the overlines denote averages in $X$.  

\begin{figure}[t]
\centering
\hspace{40pt}\includegraphics[width=0.6\textwidth]{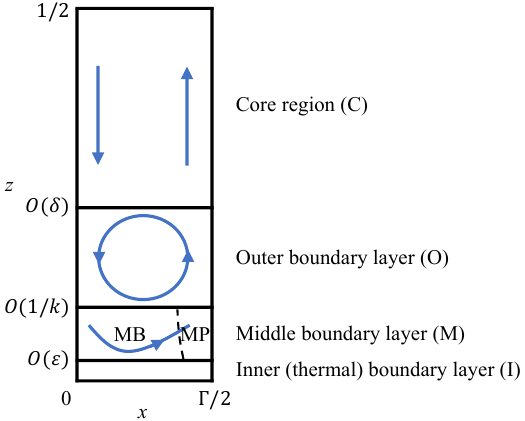}
\caption{Asymptotic structure of strongly nonlinear steady rolls whose horizontal periods scale like $\Gamma=O(Ra^{-1/4})$ as $Ra\to\infty$. The bottom half of a single roll is shown. The four vertically-stacked layers at each boundary are the core (C), outer boundary layer (O), middle boundary layer (M), and inner (thermal) boundary layer (I). The vertical thickness of each layer is $O(1)$, $O(\delta)$, $O(1/k)=\mathit{O}(\Gamma)$, and $O(\eps)$, respectively, where the scalings of $\delta$ and $\eps$ must be determined by the analysis. The middle boundary layer is divided horizontally into two regions: a bulk (MB) with homogeneous temperature and a thin plume (MP) where most of the heat flux across the layer occurs.}
\label{fig:Schematic_LocalOptimizers}
\end{figure}

Our analysis suggests that steady rolls develop an asymptotic structure with four vertically stacked regions near each boundary.  In each region, the governing PDEs~\cref{eq: steadyboussX} have different dominant terms, as detailed in the following subsections.  \Cref{fig:Schematic_LocalOptimizers} depicts these regions in the bottom half of a single roll.  In the core region (C), the dominant balance supports a heat-exchanger flow~\cref{eq:heatexchanger}.  In the outer (recirculation) boundary layer (O), the dominant balance is between advection, buoyancy, and horizontal diffusion.  In the middle boundary layer (M), advection dominates diffusion in its bulk, rendering the flow effectively inviscid there.  Most of the heat flux through the middle boundary layer is carried by a thin thermal plume, so this layer is further divided into its bulk region (MB) and plume region (MP).  Lastly, in the inner (thermal) boundary layer (I), heat flux arises from a combination of diffusion and advection, rather than diffusion being dominated by advection.  The formulas for these boundary layers refer to the bottom half of a roll, but the top half is related by symmetry.

\subsection{Core region}
\label{sec:Core}

In the core, we decompose $T(X,z)=\overline{T}(z)+\theta(X,z)$. Guided by the numerics (see figure~\ref{fig:Spectra_TPsiOmega}\emph{a,d}), we assume that $|\theta| \ll \overline{T}$ (more precisely, we require $|\partial_z \theta| \ll |\partial_z \overline{T}|$). In this region, the buoyancy torque is balanced by the horizontal ($X$) diffusion of vorticity, while vertical advection of the horizontal mean temperature $\overline{T}$ is balanced by the $X$-diffusion of the temperature anomaly $\theta$:
\begin{subequations}
\label{eq:balance_C}
\begin{align}
Ra^{-1/4}Pr^{1/2}\partial_X^2 \omega &\sim -\partial_X\theta, \label{eq:balance_C 1} \\
Ra^{1/2}\partial_X^2\psi &\sim -\omega, \label{eq:balance_C 2} \\
- \partial_X\psi\partial_z\ov{T} &\sim Ra^{-1/4}Pr^{-1/2}\partial_X^2\theta. \label{eq:balance_C 3}
\end{align}
\end{subequations}
Crucially, the prescribed aspect-ratio scaling $\Gamma = \mathit{O}(Ra^{-1/4})$ is the unique choice that enables this self-consistent balance of terms, given the assumption $\theta \to 0$ as $Ra\to\infty$. Equations~(\ref{eq:balance_C 1})--(\ref{eq:balance_C 3}) can be collapsed to $\partial_X^4\omega\sim (-\partial_z\overline{T})\omega$, implying $\partial_z\ov{T}=O(1)$ and that $\omega$ and, hence, $\psi$ and $\theta$ are separable functions of $X$ and $z$.

The leading order solutions in the core are denoted as $\omega^c$, $\psi^c$, $\theta^c$, and $\ov{T^c}$, where the $c$ superscript refers to a core field.
For the $X$-varying fields, we let
\begin{align}
\omega^c(X,z) &= \omega_c\Omega^c(X,z), &
\psi^c(X,z) &= \psi_c\Psi^c(X,z), &
\theta^c(X,z) &= \theta_c\Theta^c(X,z),
\label{eq:fields_C}
\end{align}
where the subscripted prefactors scale with $Ra$ as needed so that the capitalized fields $\Omega^c$, $\Psi^c$, and $\Theta^c$ are $O(1)$. Having already assumed $\theta\ll\overline{T}$, we let $\theta_c=Ra^s$ with $s<0$ to be determined in \cref{sec:Matching} by matching conditions. Then, the dominant balances~\cref{eq:balance_C 1,eq:balance_C 2} establish the scalings of $\omega_c$ and $\psi_c$, giving
\begin{align}
\omega_c = Ra^{s+1/4}, \quad
\psi_c = Ra^{s-1/4}, \quad
\mbox{and} \;\; \theta_c = Ra^s. 
\end{align}
These scalings are consistent with the absence of nonlinear terms from~\cref{eq: steadyboussX} in the dominant balance~\cref{eq:balance_C}, which therefore has a quasilinear structure (i.e., the $X$-varying fluctuation dynamics are linearised about the $X$-mean temperature field $\ov{T}$).  The $O(1)$ fields then satisfy
\begin{subequations}
\label{eq:leading_C}
\begin{align}
Pr^{1/2} \partial_X^2 \Omega^c &= -\partial_X\Theta^{c}, \label{eq: omegaX_C} \\
\partial_X^2 \Psi^c &= -\Omega^{c}, \label{eq: psiX_C}\\
-\partial_X\Psi^c\partial_z \overline{T^c} &= Pr^{-1/2} \partial_X^2\Theta^{c}. \label{eq: thetaX_C}
\end{align}
\end{subequations}

In the core, the horizontally averaged expression~\cref{eq:Nu(z)} for $Nu$ is dominated by its convective term, so at each $z$,
\beq
Nu\sim -Pr^{1/2}Ra^{2s+1/2}\overline{\Theta^c\partial_X\Psi^c}.
\label{eq:Nu core}
\eeq
This balance and the dominance $Nu\gg1$ imply that $-1/4<s$. From this same balance and the $z$-independence of $Nu$ one can show that 
$\partial_z\ov{T^c}$ is constant and the $X$-varying core fields depend solely on $X$. To see this, note that~\cref{eq: omegaX_C,eq: psiX_C} imply $-\ov{\Theta^c\partial_X\Psi^c} = -Pr^{1/2}\overline{\partial_X^3\Psi^c\partial_X\Psi^c} = Pr^{1/2}\overline{(\partial_X^2\Psi^c)^2} = Pr^{1/2}\overline{(\Omega^c)^2}$ after integration by parts. Given that $\overline{(\Omega^c)^2}$ is proportional to $Nu$ and that $\Omega^c$ can be expressed as a separable function of $X$ and $z$, $\Omega^c$ and, hence, from (\ref{eq:leading_C}),  $\Psi^c$ and $\Theta^c$ must therefore depend only on $X$. Finally,~\cref{eq: thetaX_C} implies that $\partial_z\ov{T^c}$ is constant. 

The $z$-independent fields solving~\cref{eq:leading_C} can be found explicitly up to a free coefficient $A$, so the leading order solution in the core is
\begin{subequations}
\label{eq:solution_C}
\begin{align}
\omega^c &= Ra^{s+1/4} \Omega^c(X) =  -Ra^{s+1/4}Pr^{-1/2}K^{-1}{A}\sin(KX),\\
\psi^c &= Ra^{s-1/4}\Psi^c(X) = -Ra^{s-1/4}Pr^{-1/2}K^{-3}{A}\sin(KX),\\
T^c &=  \overline{T^c}(z)+Ra^{s}\Theta^{c}(X) = \tfrac{1}{2} - K^4\left(z-\tfrac12\right) + Ra^s A\cos(KX),
\end{align}
\end{subequations}
where $A$ can depend on $Pr$ and $K$ but is $O(1)$ in $Ra$. This is exactly the heat-exchanger solution~\cref{eq:heatexchanger} with $A = -\tilde{A}Ra^{-s+1/4}Pr^{1/2} K^3$, which further implies $\tilde{A} = O(Ra^{s-1/4})$.  As discussed in \cref{sec:LocaMax_numerics} and also evident in \cref{eq:solution_C}, in the core region, the flow adopts a $z$-independent, single-Fourier-mode structure in the horizontal direction.  In this configuration, the vertical advection of an emergent, linearly $z$-varying horizontal-mean temperature field is balanced by horizontal diffusion.  

The heat-exchanger structure~\cref{eq:solution_C} clearly cannot extend to the boundaries, as it would violate the boundary conditions. For instance, the $z$-independent flow fails to satisfy the zero normal-flow (impenetrability) condition at the walls.  This issue is resolved by the presence of a `turnaround' region near each boundary, which is isotropic in the sense that its thickness aligns with the $O(Ra^{-1/4})$ scale of each roll's width.  The turnaround region ends up being the middle boundary layer (cf.\ \cref{fig:Schematic_LocalOptimizers}), since additional layers are needed for matching with the core and to the adjacent boundary.

\subsection{Outer boundary layer region}
\label{sec:OuterBL}

Direct asymptotic matching between the core region and the isotropic turnaround region is not possible, as explained below in~\cref{sec:MiddleBL}. However, they can be connected by an outer recirculation region. Existence of an outer layer whose thickness scales approximately as $Ra^{-0.1}$ is suggested by our numerically computed rolls, especially \cref{fig:MeanBL_TPsiOmega}($b$, $c$), but the exact scaling will be determined through asymptotic analysis. 

In all boundary layers we decompose $T=\tau +\theta$, where $\tau$ is the constant value that matches $\ov{T}$ at the edge of the core.  In the bottom boundary layers $\tau=\tfrac12(1+K^4)$. This $\tau$ value does not generally coincide with $\ov{T}$, so $\ov{\theta}$ need not vanish in the boundary layers.

The thickness of the outer boundary layer is denoted by $\delta$, so we let $z = \delta \zeta$ in the (bottom) outer layer.  In the $O(1)$ coordinates $(X,\zeta)$, equations~\cref{eq: steadyboussX} become
\begin{subequations}
\label{eq: steadyboussX_zeta}
\begin{align}
\partial_\zeta\psi\partial_X\omega - \partial_X\psi\partial_\zeta\omega  &= 
Pr^{1/2}Ra^{-1/4}\delta \left(\partial_X^2 + \delta^{-2}Ra^{-1/2}\partial_{\zeta}^2\right) \omega + \delta\partial_X \theta, \label{eq: omegaX_zeta} \\
\left(\partial_X^2 + \delta^{-2}Ra^{-1/2}\partial_{\zeta}^2\right) \psi &= -Ra^{-1/2}\omega, \label{eq: psiX_zeta}\\
\partial_\zeta\psi\partial_X \theta - \partial_X\psi\partial_\zeta \theta & = Pr^{-1/2} Ra^{-1/4} \delta \left(\partial_X^2 + \delta^{-2}Ra^{-1/2}\partial_{\zeta}^2\right) \theta. \label{eq: thetaX_zeta}
\end{align}
\end{subequations}
Since the thickness of the outer layer must be less than the $O(1)$ height of the core and greater than the $O(Ra^{-1/4})$ thickness of the middle layer defined below, meaning $Ra^{-1/4}\ll \delta \ll 1$, the diffusion terms in~\cref{eq: steadyboussX_zeta} are dominated by the $\partial_X^2$ parts.  Thus, the balance among the remaining terms---i.e., nonlinear advection, horizontal diffusion, and buoyancy torque---yields
\begin{subequations}
\label{eq:balance_O}
\begin{align}
\partial_\zeta\psi\partial_X\omega - \partial_X\psi\partial_\zeta\omega  &\sim 
Pr^{1/2}Ra^{-1/4}\delta \partial_X^2\omega + \delta\partial_X\theta, \label{eq: omegaX_zeta 2} \\
\partial_X^2\psi &\sim -Ra^{-1/2}\omega, \label{eq: psiX_zeta 2}\\
\partial_\zeta\psi\partial_X\theta - \partial_X\psi\partial_\zeta\theta &\sim Pr^{-1/2} Ra^{-1/4}\delta \partial_X^2\theta. \label{eq: thetaX_zeta 2}
\end{align}
\end{subequations}
Leading-order solutions $\omega^o$, $\psi^o$, and $\theta^o$ that satisfy~\cref{eq:balance_O} with equalities are rescaled~as
\begin{align}
\omega^o(X,\zeta) &= \omega_o\Omega^o(X,\zeta), &
\psi^o(X,\zeta) &= \psi_o\Psi^o(X,\zeta), &
\theta^o(X,\zeta) &= \theta_o\Theta^o(X,\zeta),
\end{align}
where the $o$ superscript refers to an outer boundary layer field and subscripted prefactors scale with $Ra$ as needed so that the capitalized fields $\Omega^o$, $\Psi^o$, and $\Theta^o$ are $O(1)$. 

Balancing~\cref{eq: thetaX_zeta 2} gives $\psi_o=\delta Ra^{-1/4}$, then~\cref{eq: psiX_zeta 2} gives $\omega_o=Ra^{1/2}\psi_o=\delta Ra^{1/4}$, and balancing the last term of~\cref{eq: omegaX_zeta 2} with the others gives $\theta_o=Ra^{-1/4}\omega_o=\delta$. In the outer layer, as in the core, the horizontally averaged $Nu$ expression~\cref{eq:Nu(z)} is dominated by the convective contribution,
\beq
Nu\sim -Pr^{1/2}Ra^{1/2}\delta^2\ov{\Theta^o\partial_X\Psi^o}.
\label{eq:Nu outer}
\eeq
Matching between the outer layer and the core requires the asymptotic expressions \cref{eq:Nu core,eq:Nu outer} to have the same scaling, and so
\beq
\delta=Ra^s,
\label{eq:delta}
\eeq
where we recall that $O(Ra^s)$ for some $-1/4<s<0$ has been defined as the scale of temperature fluctuations in the core. Our expansions in the (bottom) outer layer are therefore
\begin{subequations}
\label{eq:fields_O}
\begin{align}
\omega^o(X, \zeta) &= Ra^{s+1/4} \Omega^{o}(X, \zeta),\\
\psi^o(X, \zeta) &= Ra^{s-1/4} \Psi^{o}(X, \zeta), \label{eq:psi_O}\\
T^o(X, \zeta) &= \tfrac12(1 + K^4) + Ra^{s}\Theta^{o}(X, \zeta).
\end{align}
\end{subequations}
The PDEs governing the $O(1)$ rescaled fields in the outer layer are
\begin{subequations}
\label{eq:leading_O}
\begin{align}
\partial_\zeta\Psi^o\partial_X\Omega^o - \partial_X\Psi^o\partial_\zeta\Omega^o  &= 
Pr^{1/2} \partial_X^2 \Omega^o + \partial_X \Theta^o, \label{eq: omegaX_zeta_O} \\
\partial_X^2 \Psi^o &= -\Omega^o, \label{eq: psiX_zeta_O}\\
\partial_\zeta\Psi^o\partial_X \Theta^o - \partial_X\Psi^o\partial_\zeta \Theta^o & =  Pr^{-1/2} \partial_X^2 \Theta^o. \label{eq: thetaX_zeta_O}
\end{align}
\end{subequations}
We note that (\ref{eq: omegaX_zeta_O})--(\ref{eq: thetaX_zeta_O}) are the equations governing a laminar plume, but applied to flow in a domain that is bounded and periodic in the transverse (here, $X$) direction.

\subsection{Middle boundary layer region}
\label{sec:MiddleBL}

The turnaround region near each boundary is isotropic, meaning that thickness of the middle layer has the same $O(Ra^{-1/4})$ scaling as the roll width.  Accordingly, we define $z = Ra^{-1/4}Z$ in the (bottom) middle layer.  In the $O(1)$ coordinates $(X, Z)$, the equations~\cref{eq: steadyboussX} transform to
\begin{subequations}
\label{eq: steadyboussX_Z}
\begin{align}
\partial_Z\psi\partial_X\omega - \partial_X\psi\partial_Z\omega  &= 
{Pr}^{1/2}Ra^{-1/2}\left(\partial_X^2 + \partial_Z^2\right) \omega + Ra^{-1/4}\partial_X \theta, \label{eq: omegaX_Z} \\
Ra^{1/2}\left(\partial_X^2 + \partial_Z^2\right) \psi &= -\omega, \label{eq: psiX_Z}\\
\partial_Z\psi\partial_X \theta - \partial_X\psi\partial_Z \theta & = Pr^{-1/2}Ra^{-1/2} \left(\partial_X^2 + \partial_Z^2\right) \theta. \label{eq: thetaX_Z}
\end{align}
\end{subequations}
As in all bottom boundary layers, $T=\tau+\theta$ with $\tau=\tfrac12(1+K^4)$.

First, we explain why a region with this scaling cannot be asymptotically matched directly with the core. The core fields are $z$-independent, so they can only be matched to another region in which the fields have the same scalings, namely $\omega=O(Ra^{s+1/4})$, $\psi=O(Ra^{s-1/4})$, and $\theta=O(Ra^s)$, where $-1/4<s<0$.  If the fields in the turnaround region adopted these scalings, the dominant balance in~\cref{eq: steadyboussX_Z} would occur solely among the $O(Ra^{2s})$ left-hand terms, as the other terms would scale as $O(Ra^{s-1/4})$.  In contrast, the dominant balance in the core involves only the right-hand terms of the same equations.  Since there would be no common dominant terms in both the turnaround region and the core, these regions cannot be directly matched.  This issue is resolved by identifying the turnaround region as the middle layer. The outer layer introduced in \cref{sec:OuterBL} ensures proper matching between this middle layer and the core is achievable.

\begin{figure}[t]
\centering
\includegraphics[width=0.98\textwidth]{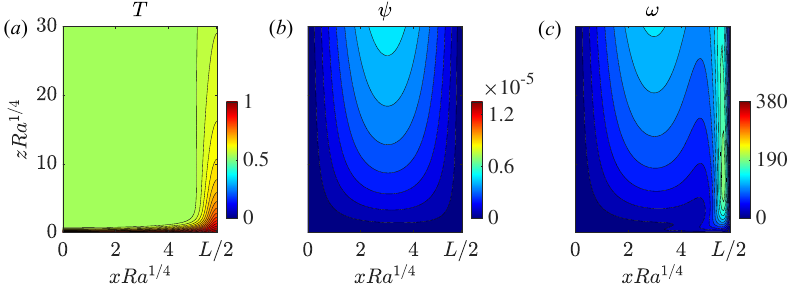}
\caption{Contours of ($a$) $T$, ($b$) $\psi$ and ($c$) $\omega$ for steady roll of locally $Nu$-maximizing period $\Gamma^*_{loc}\approx 0.00117$ at $Ra=10^{16}$ and $Pr=0.1$. Both coordinates are rescaled by $Ra^{1/4}$ as in the middle boundary layer, although the axis scales are unequal.}
\label{fig:T_Psi_Omega_MiddleBL}
\end{figure}

The middle layer is further divided horizontally into two regions with distinct scalings, which are supported by our numerics. \Cref{fig:T_Psi_Omega_MiddleBL} illustrates a portion of a numerically computed roll near the bottom boundary. A thermal plume is evident at the right edge of the temperature field in \cref{fig:T_Psi_Omega_MiddleBL}($a$), where hot fluid rises upward. The plume region of the middle layer (MP) is an asymptotically thin zone near the right edge of the roll ($x=\Gamma/2$ or $X = xRa^{1/4} = L/2$), characterized by a relatively large temperature variation. While the precise scaling of this region remains to be determined, it is clear that $L/2 - X \ll 1$ within the plume as $Ra\to\infty$. The majority of heat flux across the middle layer occurs within this narrow region.  The bulk region of the middle layer (MB), characterized by a nearly constant temperature, as indicated by the numerics (cf.~\cref{fig:T_Psi_Omega_MiddleBL}$a$), encompasses all $X$ values except those asymptotically close to $L/2$, where the effects of the plume become dominant.

\subsubsection{Bulk region of the middle boundary layer}

In the bulk region of the middle boundary layer, advection dominates diffusion, so
\begin{subequations}
\label{eq:balance_MB}
\begin{align}
\partial_Z\psi\partial_X\omega &\sim \partial_X\psi\partial_Z\omega, \label{eq: omegaX_Z 2} \\
Ra^{1/2}\left(\partial_X^2 + \partial_Z^2\right) \psi &\sim -\omega, \label{eq: psiX_Z 2}\\
\partial_Z\psi\partial_X\theta &\sim \partial_X\psi\partial_Z\theta. \label{eq: thetaX_Z 2}
\end{align}
\end{subequations}
Although it is not yet clear whether the buoyancy torque in \cref{eq: omegaX_Z} is of the same order as the advection terms, subsequent matching between different regions suggests that the buoyancy torque is, in fact, a smaller term.  The leading order solutions $\omega^{mb}$, $\psi^{mb}$, and $\theta^{mb}$ are rescaled as
\begin{align}
\omega^{mb} &= \omega_{mb}\Omega^{mb}(X,Z), &
\psi^{mb} &= \psi_{mb}\Psi^{mb}(X,Z), &
\theta^{mb} &= \theta_{mb}\Theta^{mb}(X,Z),
\end{align}
where the $mb$ superscript refers to a bulk region field within the middle boundary layer, and the subscripted prefactors scale with $Ra$ appropriately so that the capitalized fields $\Omega^{mb}$, $\Psi^{mb}$, and $\Theta^{mb}$ are $O(1)$.  Numerical results indicate that the temperature in this bulk region remains nearly constant and comparable to its characteristic value in the outer boundary layer, implying $\theta_{mb}=Ra^s$.  To determine the scaling of the stream function, we define $\psi_{mb}=Ra^\gamma$, where $\gamma$ will be identified through matching conditions in subsequent analysis.  Using \cref{eq: psiX_Z 2}, the scaling of vorticity follows as $\omega_{mb}=Ra^{\gamma+1/2}$.  Consequently, the expansions for the fields in the bulk region are
\begin{subequations}
\label{eq:fields_MB}
\begin{align}
\omega^{mb}(X, Z) &= Ra^{\gamma + 1/2} \Omega^{mb}(X, Z),\\
\psi^{mb}(X, Z) &= Ra^{\gamma} \Psi^{mb}(X, Z), \label{eq:psi_MB}\\
T^{mb}(X, Z) &=  \tfrac12(1 + K^4) + Ra^{s}\Theta^{mb}(X, Z).
\end{align}
\end{subequations}
Provided that $-1/2<\gamma$ and the buoyancy torque is a smaller term (as can be confirmed \emph{a posteriori}), the dominant balance of~\cref{eq: steadyboussX_Z} indeed reduces to~\cref{eq:balance_MB}.  Since $\omega^{mb}$, $\psi^{mb}$, and $\theta^{mb}$ solve~\cref{eq:balance_MB} with equality, their $O(1)$ rescalings satisfy the following equations
\begin{subequations}
\label{eq:leading_MB}
\begin{align}
\partial_Z\Psi^{mb}\partial_X\Omega^{mb} - \partial_X\Psi^{mb}\partial_Z\Omega^{mb}  &= 0, \label{eq: omegaX_Z_MB} \\
\left(\partial_X^2 + \partial_Z^2\right) \Psi^{mb} &= -\Omega^{mb}, \label{eq: psiX_Z_MB}\\
\partial_Z\Psi^{mb}\partial_X \Theta^{mb} - \partial_X\Psi^{mb}\partial_Z \Theta^{mb} & =   0. \label{eq: thetaX_Z_MB}
\end{align}
\end{subequations}

The growth of the velocity field as $Z\to\infty$ is dictated by matching the bulk region of the middle layer with the outer layer.  In particular, the limits of $Ra^{\gamma} \Psi^{mb}(X, Z)$ as $Z\to\infty$ and $Ra^{s - 1/4}\Psi^o(X, \zeta)$ as $\zeta\to 0^+$ must exhibit the same dependence on $Ra$.  Following \cite{Deguchi2023}, we assume that $\Psi^{mb}$ grows algebraically as $Z^\lambda$ for some $\lambda>0$.  Then, the $Z\to\infty$ limit scales as $O(Ra^{\gamma+\lambda(s+1/4)})$ since $Z = Ra^{s + 1/4}\zeta$.  Similarly, the second limit scales as $O(Ra^{s-1/4})$.  Equating the exponents yields
\begin{equation}
\label{eq:matching_MbO}
\lambda = \frac{s - 1/4 - \gamma}{s + 1/4}.
\end{equation}
This formula gives the value of the growth rate $\lambda$ once $s$ and $\gamma$ are determined through subsequent matching.

As in \cite{Deguchi2023}, matching between the middle and outer layers more specifically requires that $\Psi^{mb}(X,Z)\sim \Psi^o(X,\zeta\to 0^+) Z^\lambda$ as $Z\to \infty$. If it is assumed that the bulk of the middle layer carries the heat flux (with the heat flux being dominated by the advective contribution there), then the $Z$-independence of the heat flux would require that $\Theta^{mb}(X,Z)\sim \Theta^o(X,\zeta\to 0^+) Z^{-\lambda}$ as $Z\to \infty$; again, see \cite{Deguchi2023}. Noting that the formal solution of (\ref{eq: thetaX_Z_MB}) is $\Theta^{mb} = \Theta^{mb}(\Psi^{mb})$, however, this far-field behaviour would necessitate that $\Theta^{mb}(\Psi^{mb})  = C^{mb}/\Psi^{mb}$, for some constant $C^{mb}$. This dependency then would imply the divergence of the temperature fluctuation $\Theta^{mb}$ as $\Psi^{mb}\to 0$, i.e., as $X\to 0^+$, as $Z\to 0^+$ and as $X\to (L/2)^-$, behaviour that is unmatchable and not evident in our numerical solutions at large $Ra$. Instead, motivated by our numerics, we introduce a thin plume region along the right-hand edge of the middle boundary layer that carries the heat flux, as discussed in the following subsection. Crucially, the introduction of the plume region relaxes the constraint on the functional form of $\Theta^{mb}(\Psi^{mb})$, allowing matchable behaviour that accords with our numerical results. Finally, as noted above, the buoyancy torque is subdominant in the bulk of the middle layer, in contrast to \cite{Deguchi2023}. The extent of the outer boundary layer in $\zeta$ is semi-infinite, allowing diffusion to suppress the $X$-variation of the temperature fluctuation $\Theta^o(X,\zeta)$ in the downwelling leg of the convection cell as the bulk of the middle layer is approached.  

\subsubsection{Plume region of the middle boundary layer}

The width of the plume region within the middle layer is $O(Ra^{-\beta})$, where the constant $\beta>0$ is yet to be determined.  Let $X = {L}/{2} + Ra^{-\beta}\chi$, so the $\chi$ coordinate remains $O(1)$ and is negative in the plume region for the single-roll structure depicted in \cref{fig:Schematic_LocalOptimizers}.  In the $(\chi,Z)$ coordinates, equations~\cref{eq: steadyboussX_Z} become
\begin{subequations}
\label{eq: steadyboussCHI_Z}
\begin{align}
\partial_Z\psi\partial_\chi\omega - \partial_\chi\psi\partial_Z\omega  &= 
{Pr}^{1/2}Ra^{\beta-1/2}\left(\partial_\chi^2 + Ra^{-2\beta}\partial_Z^2\right) \omega + Ra^{-1/4}\partial_\chi\theta, \label{eq: omegaCHI_Z} \\
Ra^{2\beta+1/2}\left(\partial_\chi^2 + Ra^{-2\beta}\partial_Z^2\right) \psi &= -\omega, \label{eq: psiCHI_Z}\\
\partial_Z\psi\partial_\chi \theta - \partial_\chi\psi\partial_Z \theta &= Pr^{-1/2}Ra^{\beta-1/2} \left(\partial_\chi^2 + Ra^{-2\beta}\partial_Z^2\right)\theta. \label{eq: thetaCHI_Z}
\end{align}
\end{subequations}
Within the narrow plume region, the second partial vertical derivatives can be neglected in the Laplacian operators, while all other terms in the governing equations are retained, yielding the following approximate system:
\begin{subequations}
\label{eq:balanceMP}
\begin{align}
\partial_Z\psi\partial_\chi\omega - \partial_\chi\psi\partial_Z\omega  &\sim 
{Pr}^{1/2}Ra^{\beta-1/2}\partial_\chi^2\omega + Ra^{-1/4}\partial_\chi\theta,\\
Ra^{2\beta+1/2}\partial_\chi^2\psi &\sim -\omega,\\
\partial_Z\psi\partial_\chi \theta - \partial_\chi\psi\partial_Z\theta & \sim Pr^{-1/2}Ra^{\beta-1/2}\partial_\chi^2\theta.
\end{align}
\end{subequations}
The fields that solve~\cref{eq:balanceMP} with equalities are represented as
\begin{align}
\omega^{mp} &= \omega_{mp}\Omega^{mp}(\chi,Z), &
\psi^{mp} &= \psi_{mp}\Psi^{mp}(\chi,Z), &
\theta^{mp} &= \theta_{mp}\Theta^{mp}(\chi,Z),
\end{align}
where the $mp$ superscript refers to a plume region field within the middle boundary layer, and the subscripted prefactors scale with $Ra$ appropriately so that the capitalized fields $\Omega^{mp}$, $\Psi^{mp}$, and $\Theta^{mp}$ are $O(1)$.  Balancing all terms in~\cref{eq:balanceMP} requires $\psi_{mp}=Ra^{\beta-1/2}$, $\omega_{mp}=Ra^{3\beta}$ and $\theta_{mp}=Ra^{4\beta-1/4}$. Thus, our expansions in the plume region are
\begin{subequations}
\label{eq:fields_MP}
\begin{align}
\omega^{mp}(\chi, Z) &= Ra^{3\beta} \Omega^{mp}(\chi, Z),\\
\psi^{mp}(\chi, Z) &= Ra^{\beta - 1/2} \Psi^{mp}(\chi, Z), \label{eq:psi_MP}\\
T^{mp}(\chi, Z) &=  \tfrac12(1+K^4) + Ra^{4\beta - 1/4}\Theta^{mp}(\chi, Z).
\end{align}
\end{subequations}
For these to satisfy~\cref{eq:balanceMP} with equalities, the $O(1)$ fields satisfy
\begin{subequations}
\label{eq:leading_MP}
\begin{align}
\partial_Z\Psi^{mp}\partial_\chi\Omega^{mp} - \partial_\chi\Psi^{mp}\partial_Z\Omega^{mp}  &= 
{Pr}^{1/2} \partial_\chi^2 \Omega^{mp} + \partial_\chi \Theta^{mp}, \label{eq: omegaCHI_Z_MP} \\
\partial_\chi^2 \Psi^{mp}\ &= -\Omega^{mp}, \label{eq: psiCHI_Z_MP}\\
\partial_Z\Psi^{mp}\partial_\chi \Theta^{mp} - \partial_\chi\Psi^{mp}\partial_Z \Theta^{mp} & =   Pr^{-1/2} \partial_\chi^2 \Theta^{mp}. \label{eq: thetaCHI_Z_MP}
\end{align}
\end{subequations}

In the horizontally averaged expression~\cref{eq:Nu(z)} for $Nu$, we assume and confirm \emph{a posteriori} that the convective part is not only dominant, as it is in the core and outer layer, but also primarily driven by the plume region.  Thus, at each height $z$ in the (bottom) middle layer,
\begin{align}
\label{eq:Nu(z)_MBL}
Nu &\sim - Pr^{1/2}Ra^{3/4} \left(\frac{2}{L}\int_{\text{MP}}\theta\partial_X\psi\,{\rm d}X\right) = - Pr^{1/2}Ra^{5\beta}\left(\frac{2}{L}\int_{-\infty}^0\Theta^{mp}\partial_\chi\Psi^{mp}\,{\rm d}\chi\right),
\end{align}
where the first integral is from the plume edge at $X=L/2-O(Ra^{-\beta})$ to the roll boundary at $X=L/2$. The heat transport across the plume must match that across the outer layer and the core, so it must scale with $Ra$ the same way as in~\cref{eq:Nu core}. This means that $5\beta=2s+1/2$, so 
\begin{align}
\label{eq:Nu(z)_MpBL}
\beta = \frac{2}{5}s+\frac{1}{10}.
\end{align}

Matching stream functions between the plume and bulk regions of the middle boundary layer establishes a relationship between the $O(Ra^{-\beta})$ width of the plume and the $O(Ra^\gamma)$ magnitude of $\psi$ in the bulk.  Specifically, the limits of $Ra^{\beta - 1/2}\Psi^{mp}(\chi, Z)$ as $\chi\to-\infty$ and $Ra^{\gamma} \Psi^{mb}(X, Z)$ as $X\to (L/2)^-$ must scale consistently with $Ra$.  Assume that $\Psi^{mp}$ grows as $O(\chi^{d})$ in the limit $\chi\to-\infty$, with the exponent $d$ to be determined.  Matching the scaling of $\psi$ requires $Ra^{(1+d)\beta - 1/2} = Ra^\gamma$, leading to the relation
\begin{equation}
\label{eq:matching_MbMp}
 \gamma = (1+d)\beta - \frac{1}{2}.
 \end{equation}

\subsection{Inner boundary layer region}
\label{sec:InnerBL}

An inner (thermal) boundary layer is required to satisfy the temperature and stress-free (or no-slip, in another configuration) conditions imposed at the boundary, as further evidenced by the numerically computed rolls shown in \cref{fig:Spectra_TPsiOmega}.  The thickness of this inner boundary layer is denoted by $\eps$, and we set $z = \eps\eta$ within the layer.  In the $O(1)$ $(X,\eta)$ coordinates, the governing PDEs~\cref{eq: steadyboussX} transform to
\begin{subequations}
\label{eq: steadyboussX_eta}
\begin{align}
\partial_\eta\psi\partial_X\omega - \partial_X\psi\partial_\eta\omega  &= 
Pr^{1/2}Ra^{-3/4}\eps^{-1} \left(\partial_{\eta}^2 + \eps^2 Ra^{1/2}\partial_X^2\right) \omega + \eps\partial_X \theta, \label{eq: omegaX_eta} \\
\left(\partial_\eta^2 + \eps^{2}Ra^{1/2}\partial_X^2\right) \psi &= -\eps^{2}\omega, \label{eq: psiX_eta}\\
\partial_\eta\psi\partial_X\theta - \partial_X\psi\partial_\eta\theta & = Pr^{-1/2}Ra^{-3/4}\eps^{-1} \left(\partial_{\eta}^2 + \eps^2 Ra^{1/2}\partial_X^2\right)\theta. \label{eq: thetaX_eta}
\end{align}
\end{subequations}
As for all the boundary layers, we write $T=\tau+\theta$ with $\tau$ being constant.  Since the inner boundary layer is thinner than the middle layer, we have $\eps \ll Ra^{-1/4}$. Consequently, within the inner boundary layer, the vertical derivatives in each Laplacian operator dominate horizontal derivatives, yielding
\begin{subequations}
\label{eq:balance_I}
\begin{align}
\partial_\eta\psi\partial_X\omega - \partial_X\psi\partial_\eta\omega &\sim 
Pr^{1/2}Ra^{-3/4}\eps^{-1}\partial_{\eta}^2\omega + \eps\partial_X \theta, \label{eq:balance_I 1} \\
\partial_\eta^2\psi &\sim -\eps^{2}\omega, \label{eq:balance_I 2} \\
\partial_\eta\psi\partial_X\theta - \partial_X\psi\partial_\eta\theta &\sim Pr^{-1/2}Ra^{-3/4}\eps^{-1} \partial_{\eta}^2\theta. \label{eq:balance_I 3}
\end{align}
\end{subequations}
The balance~\cref{eq:balance_I 2} clearly is necessary in the no-slip scenario. Unlike the analysis of $\mathit{O}(1)$ aspect-ratio steady stress-free convection by \cite{ChiniCox2009}, in which the thermal boundary layers were found to be passive and the buoyancy torque subdominant,  here the buoyancy torque arises in the leading-order balance \cref{eq:balance_I 1}. Inclusion of this effect in turn necessitates the retention of the right-hand side of \cref{eq:balance_I 2} even in the stress-free case. Physically, the buoyancy torque renders even the leading-order wall-parallel flow within the inner layer rotational; this is to be contrasted with the the plug-like flow (i.e., with zero shear) that is realised in the stress-free case at $\mathit{O}(1)$ aspect-ratio.  Collectively, these considerations support the validity of our asymptotic construction for both stress-free and no-slip steady convection with $\Gamma = \mathit{O}(Ra^{-1/4})$.

The leading order solutions $\omega^{i}$, $\psi^{i}$, and $\theta^{i}$ that satisfy \cref{eq:balance_I} with equalities are rescaled as
\begin{align}
\omega^{i}(X,\eta) &= \omega_{i}\Omega^{i}(X,\eta), &
\psi^{i}(X,\eta) &= \psi_{i}\Psi^{i}(X,\eta), &
\theta^{i}(X,\eta) &= \theta_{i}\Theta^{i}(X,\eta),
\end{align}
where the $i$ superscript refers to an inner boundary layer field, and the subscripted prefactors scale with $Ra$ as necessary so that the capitalized fields $\Omega^{i}$, $\Psi^{i}$, and $\Theta^{i}$ remain $O(1)$.  
Balancing the terms in~\cref{eq:balance_I 2,eq:balance_I 3} yields the scalings $\psi_i=\eps^{-1}Ra^{-3/4}$ and $\omega_i=\eps^{-3}Ra^{-3/4}$,  which also ensure that the nonlinear advection terms and vertical diffusion balance in~\cref{eq:balance_I 1}.  Meanwhile, $\theta_i=1$, since the change in $T$ across the inner boundary layer is $O(1)$, and $\theta$ differs from $T$ only by a constant.  Balancing vertical diffusion with the buoyancy torque in~\cref{eq:balance_I 1} gives the relation $\eps^{-4}Ra^{-3/2} = \eps$, which leads to the scaling
 \begin{align}\label{eq:epsilon_I}
\eps  = Ra^{-3/10}.
\end{align}
This result indicates that heat transport in these roll solutions is controlled by the inner (thermal) boundary layer.  Away from the bottom wall, heat is mainly transported upward by a passive plume that emerges from the middle boundary layer. With these scalings, both the conductive and convective terms in the horizontally averaged $Nu$ expression~\cref{eq:Nu(z)} contribute at leading order within the inner boundary layer, unlike in all other regions where the convective flux dominates.  

The expansions for the inner boundary layer fields therefore are
\begin{subequations}
\label{eq:fields_I}
\begin{align}
\omega^i(X, \eta) &= \eps^{-3} Ra^{-3/4} \Omega^{i}(X, \eta),\\
\psi^i(X, \eta) &= \eps^{-1} Ra^{-3/4} \Psi^{i}(X, \eta), \label{eq:psi_I}\\
T^i(X, \eta) &=  \tfrac12(1 + K^4) + \Theta^{i}(X, \eta). \label{eq:T_I}
\end{align}
\end{subequations}
These expansions satisfy \cref{eq:balance_I} with equalities, such that the $O(1)$ capitalized fields are governed by
\begin{subequations}
\label{eq:leading_I}
\begin{align}
\partial_\eta\Psi^i\partial_X\Omega^i - \partial_X\Psi^i\partial_\eta\Omega^i  &= 
Pr^{1/2}\partial_{\eta}^2 \Omega^i + \partial_X \Theta^i, \label{eq: omegaX_eta_I}\\
\partial_\eta^2 \Psi^i &= -\Omega^i, \label{eq: psiX_eta_I}\\
\partial_\eta\Psi^i\partial_X \Theta^i - \partial_X\Psi^i\partial_\eta \Theta^i  & =  Pr^{-1/2} \partial_{\eta}^2 \Theta^i. \label{eq: thetaX_eta_I}
\end{align}
\end{subequations}

Matching $\psi$ between the inner boundary layer region and the bulk region of the middle boundary layer requires $\displaystyle\lim_{\eta\to \infty} \psi_i\Psi^i(X, \eta) = \displaystyle\lim_{Z\to 0^+} Ra^{\gamma} \Psi^{mb}(X, Z)$. As in other boundary layer analyses \citep{Schlichting2016}, we assume $\Psi^i\sim H(X)\eta$ as $\eta\to \infty$ and $\Psi^{mb}\sim H(X)Z$ as $Z\to 0^+$ for some function $H(X)$. Since $\eta =  \eps^{-1} Ra^{-1/4}Z$, matching these limits requires $\eps^{-2}=O(Ra^{1+\gamma})$.

As discussed above, the temperature change across the inner boundary layer is $O(1)$, and the horizontally averaged expression \cref{eq:Nu(z)} for $Nu$ has a leading-order contribution from its conductive term, $-\partial_z\ov{T}$.  Consequently, the inner boundary layer exhibits an $O(\eps^{-1})$ slope, leading to $Nu = O(\eps^{-1})$.  This heat flux must balance the $Nu = O(Ra^{2s+1/2})$ scaling in the core, implying $\eps=O(Ra^{-2s-1/2})$.  Furthermore, matching between the inner boundary layer and the bulk region of the middle boundary layer requires $\eps^{-2} = O(Ra^{1+\gamma})$, as explained in the preceding paragraph.  Thus, we obtain
\begin{equation}
\label{eq:matching_MI}
\gamma = 4s.
\end{equation}

\subsection{Final scalings}
\label{sec:Matching}

The scalings utilized for the five regions in \cref{sec:Core,sec:OuterBL,sec:MiddleBL,sec:InnerBL} have been expressed in terms of the exponents $s$, $\beta$, and $\gamma$.  Specifically, the scalings of the inner and outer boundary layer thicknesses, $\eps=O(Ra^{-2s-1/2})$ and $\delta=O(Ra^s)$, are determined by requiring that the horizontally averaged heat flux remains invariant with height.  This constancy constraint, combined with the necessity of matching field magnitudes across adjacent regions, yields the relations~\cref{eq:Nu(z)_MpBL}, \cref{eq:matching_MbMp}, and \cref{eq:matching_MI} between $s$, $\beta$, and $\gamma$.  Additionally, balancing vertical diffusion with both the advection terms and the buoyancy torque within the inner boundary layer establishes that $\eps = Ra^{-3/10}$~\cref{eq:epsilon_I}.  Together, these relations uniquely determine the exponents as
\begin{align}
\label{eq:scalingvalues}
s = -\frac{1}{10}, \quad\quad \beta = \frac{3}{50}, \quad\quad \mbox{and} \quad\quad \gamma = -\frac{2}{5}.
\end{align}
Consequently, the thickness of the outer boundary layer scales as
\begin{align}
\label{eq:layerthickness}
\delta &= Ra^{-{1}/{10}},
\end{align}
and the expressions~\cref{eq:matching_MbO} and \cref{{eq:matching_MbMp}} for the growth rates of stream function in the MB and MP regions provide 
\begin{align}
\lambda = 1/3 \quad\quad \mbox{and}\quad\quad d = 2/3.
\end{align}
The quantitative scalings for all five regions, which are shown schematically in \cref{fig:Schematic_LocalOptimizers}, are summarized in \cref{tab:Order}.  For each region, \cref{tab:Order} specifies its size in $(X, z)$ coordinates and the magnitudes of $\omega$, $\psi$ and $\theta$.

\begin{table}[t!]
 \begin{center}
  \begin{tabular}{lccccc}
  \hline
                 & \quad\quad$\omega$\quad\quad   & \quad\quad$\psi$\quad\quad   	&   \quad\quad$\theta$\quad\quad  & \quad\quad${L}/{2} - X$\quad\quad &  $z$\\[3pt]
       C	& $Ra^{\frac{3}{20}}$		  	& $Ra^{-\frac{7}{20}}$			& $Ra^{-\frac{1}{10}}$ 			   	& $O(1)$	& $O(1)$\\[3pt]
       O	& $Ra^{\frac{3}{20}}$			& $Ra^{-\frac{7}{20}}$			& $Ra^{-\frac{1}{10}}$ 				& $O(1)$	& $Ra^{-\frac{1}{10}}$\\[3pt]
       MB	& $Ra^{\frac{1}{10}}$			& $Ra^{-\frac{2}{5}}$				& $Ra^{-\frac{1}{10}}$ 				& $O(1)$	& $Ra^{-\frac{1}{4}}$\\[3pt]
       MP	& $Ra^{\frac{9}{50}}$			& $Ra^{-\frac{11}{25}}$			& $Ra^{-\frac{1}{100}}$ 				& $Ra^{-\frac{3}{50}}$	& $Ra^{-\frac{1}{4}}$\\[3pt]
       I		& $Ra^{\frac{3}{20}}$			& $Ra^{-\frac{9}{20}}$			& $O(1)$ 							& $O(1)$	& $Ra^{-\frac{3}{10}}$\\[3pt]
       \hline
\end{tabular}
\caption{Orders of magnitude of $\omega,\psi,\theta$ and region sizes in $(X, z)$ coordinates for the five asymptotic regions in the bottom half of a strongly nonlinear roll with $\Gamma = O(Ra^{-1/4})$.  The location of each region is illustrated in \cref{fig:Schematic_LocalOptimizers}.  The asymptotic expansions of $\omega,\psi,\theta$ in the five regions are given by~\cref{eq:solution_C,eq:fields_O,eq:fields_MB,eq:fields_MP,eq:fields_I}, and these show how $\theta$ differs from total temperature $T$ in each region.}
\label{tab:Order}
\end{center}
\end{table}

Having determined all exponents by matching, the scaling of $Nu$ can be deduced from any of its asymptotic expressions in different regions, such as $Nu=O(\eps^{-1})$ in the inner layer or $Nu=O(Ra^{2s+1/2})$ in the core.  Each expression yields the same result:
\begin{equation}
\label{eq:Nu_scaling}
Nu \sim c_n(Pr, K) Ra^{3/10},
\end{equation}
where $c_n(Pr, K)$ is a prefactor that depends on $Pr$ and $K$.  The dominant contribution to the Reynolds number~\cref{eq: Re def} arises from the core region.  Using $Re\sim Ra^{s+1/2}Pr^{-1/2} \langle(\partial_X\Psi^c)^2\rangle^{1/2}$, we find that
\begin{equation}
\label{eq:Re_scaling}
Re \sim c_r(Pr, K) Pr^{-1}Ra^{2/5}.
\end{equation}
The prefactor $c_r(Pr,K)$ has been defined with one $Pr^{-1}$ factor excluded to be consistent with the $c_r$ used in \citet{Wen2020JFM} for $k=O(1)$ rolls, where this $c_r$ is independent of $Pr$. The scalings $Nu =O(Ra^{3/10})$ and $Re = O(Ra^{2/5})$ derived here differ slightly from the predictions of \citet{Deguchi2023}, namely $Nu =O(Ra^{3/10}(\ln Ra)^{1/5})$ and $Re = O(Ra^{2/5}(\ln Ra)^{1/10})$, which omit the plume region in the middle boundary layer.

The prefactors $c_n$ and $c_r$ can be related to each other using the flow~\cref{eq:solution_C} in the core, which is determined up to its unknown amplitude $A(Pr,K)$.  Specifically, the core expression~\cref{eq:Nu core} for $Nu$ implies $c_n\sim 2^{-1} A^2K^{-2}$, while the expression~\cref{eq: Re def} for $Re$ in the core, i.e., $Re\sim Ra^{s+1/2}Pr^{-1/2} \langle(\partial_X\Psi^c)^2\rangle^{1/2}$, implies $c_r\sim2^{-1/2}|A|K^{-2}$.  Combining these results yields the relation
\beq
c_r\sim K^{-1}c_n^{1/2},
\eeq
which holds for all fixed $(Pr, K)$.  To determine the values of $c_n$ and $c_r$ \emph{a priori} for given $(Pr, K)$, it would be necessary to solve the leading order PDEs~\cref{eq:leading_O,eq:leading_MB,eq:leading_MP,eq:leading_I} in each region, subject to matching conditions between adjacent regions and bounding conditions at the domain boundaries.  By maximizing $c_n$ over $K$, one could find the asymptotic prefactor for $\Gamma^*_{loc}=O(Ra^{-1/4})$, at least when $Pr$ is sufficiently large for such a maximum to exist.  However, solution and matching of the PDEs in each region likely must be done numerically, and this is left for future work.

\subsection{Verification of asymptotic analysis with numerical results}
\label{sec:Verification}

The numerically computed rolls with locally $Nu$-maximizing periods $\Gamma^*_{loc}$, previously reported in \cref{sec:LocaMax_numerics} for $Ra$ up to $10^{16}$, sufficed to guide the matched asymptotic analysis presented above.  To further test the predictions of the asymptotic analysis, additional numerical computations have been conducted for $Ra$ up to $10^{19}$.  In these extended computations, the wavenumber $k$ is set to scale precisely with $Ra^{1/4}$, whereas the earlier results for $k^*_{loc}$ from \cref{sec:LocaMax_numerics} only approximately suggested this scaling.  Specifically, for $Pr=0.1$, 1, and 10, we fix $K=k/Ra^{1/4}=0.5361$, 0.5597, and 0.5755, respectively, to approximate the asymptotic values of $k^*_{loc}/Ra^{1/4}$, as indicated in \cref{fig:NuLocPrfixed}.

\begin{figure}[t]
\centering
\includegraphics[width=0.98\textwidth]{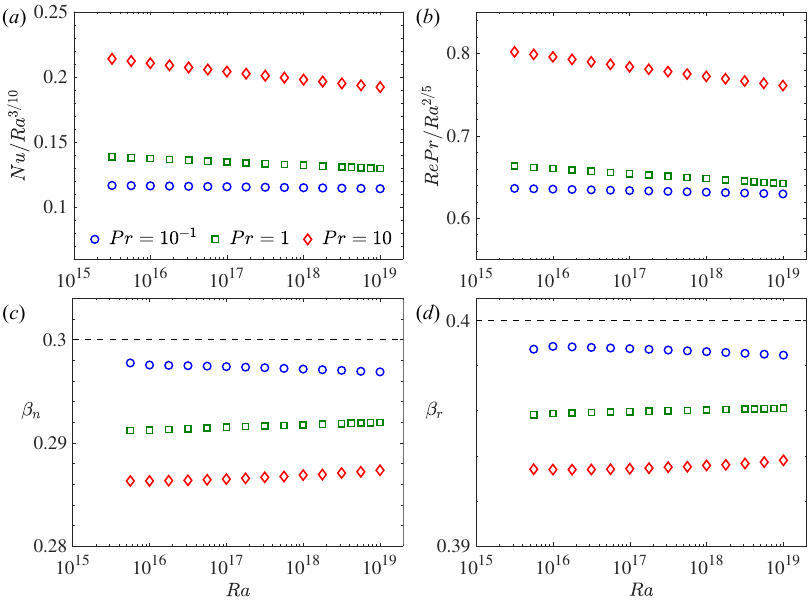}
\caption{The $Ra$-dependence of ($a$) $Nu$ and ($b$) $Re$, compensated by the scalings of our asymptotic predictions, along with the best-fit local scaling exponents ($c$) $\beta_n$ and ($d$) $\beta_r$ for high-wavenumber steady rolls with fixed $k/Ra^{1/4}$.  The values $k/Ra^{1/4}=0.5361$, 0.5597 and 0.5755 correspond to $Pr=0.1$, 1 and 10, respectively, as explained in the text.  Dashed lines in ($c$) and ($d$) indicate the scaling exponents used to compensate the plots in ($a$) and ($b$).}
\label{fig:Nu_Re_Ra_GammaOneQuarter}
\end{figure}

For rolls with fixed $k/Ra^{1/4}$, \cref{fig:Nu_Re_Ra_GammaOneQuarter} illustrates the dependence of $Nu$ and $Re$ on $Ra$.  The top plots are compensated by the predicted scalings~\cref{eq:Nu_scaling,eq:Re_scaling}, which would appear as horizontal lines if the predictions hold exactly.  The bottom plots show best-fit local scaling exponents for $Nu\propto Ra^{\beta_n}$ and $Re\propto Ra^{\beta_r}$, with dashed lines indicating the asymptotic predictions.  For $Pr=0.1$, the computed exponents $\beta_n$ and $\beta_r$ are approximately 0.2970 and 0.3985, respectively, aligning closely with the asymptotic predictions 3/10 and 2/5.  For $Pr=1$ and 10, the agreement is slightly less precise, with discrepancies within about 0.01 of the predicted exponents, but the values are trending towards the predictions as $Ra$ increases.  While we cannot definitively conclude whether all local exponents would converge more closely to the predicted values at $Ra \gg10^{19}$, the detailed analysis of the numerically computed rolls within each asymptotic region presented below provides additional support for our theoretical predictions.

\begin{figure}[t]
\centering
\includegraphics[width=0.98\textwidth]{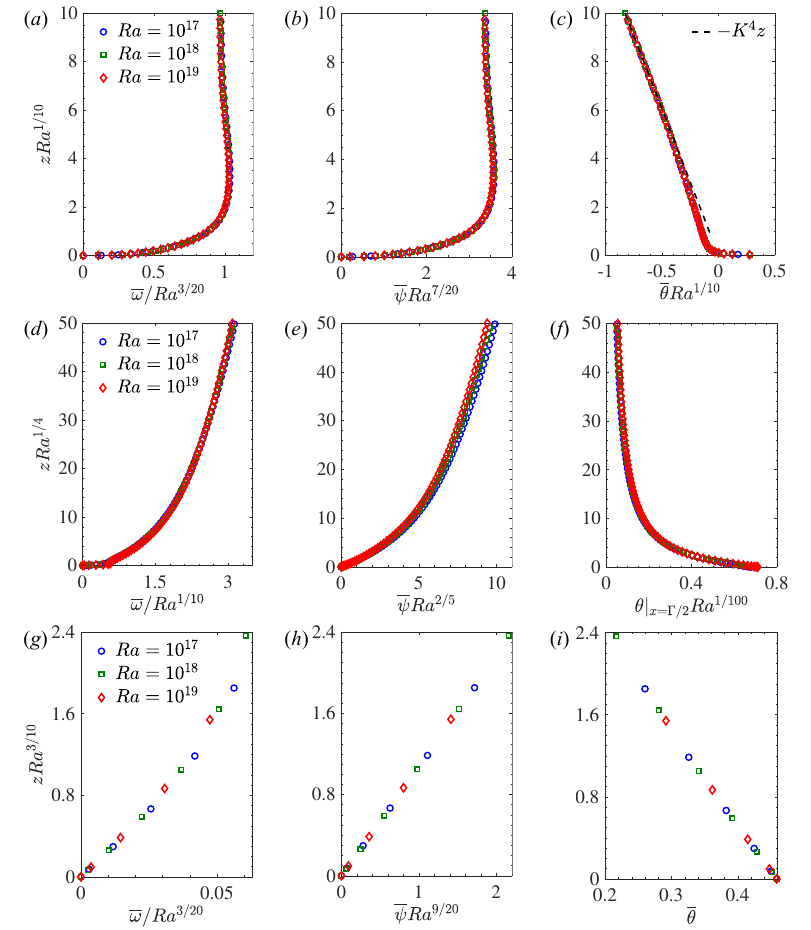}
\caption{Spatial structure of $\omega$, $\psi$ and $\theta$ near the lower boundary for a single counterclockwise roll with $Pr=0.1$ and $k/Ra^{1/4} = 0.5361$ at $Ra=10^{17}$, $10^{18}$ and $10^{19}$. Following the scaling predictions summarized in \cref{tab:Order}, variables are rescaled to collapse different $Ra$ values in the outer boundary layer (top row), middle layer (middle row) and inner layer (bottom row). In all boundary layers, $\theta = T - \tfrac12(1+K^4)$. Overlines denote horizontal averages over a single roll. For clarity, only a subset of points are plotted in the top row.}
\label{fig:Profiles_variousBLs}
\end{figure}

To test our scaling predictions for the three boundary layers, we examine the collapse of vertical profiles at different $Ra$ values when rescaling the $z$ coordinate and dependent variables according to \cref{tab:Order}.  We focus on numerically computed rolls with $(Pr,k/Ra^{1/4})=(0.1,0.5361)$, as their $Nu$ and $Re$ values in \cref{fig:Nu_Re_Ra_GammaOneQuarter} align more closely with asymptotic behaviour compared to rolls with $Pr=1$ or 10.  The top, middle, and bottom rows of \cref{fig:Profiles_variousBLs} show rescaled profiles for the outer, middle, and inner boundary layers, respectively.  In all cases, the curves at $Ra=10^{17}$, $10^{18}$ and $10^{19}$ collapse excellently, validating the predicted scalings.

The top row of \cref{fig:Profiles_variousBLs} illustrates the scaling for the outer boundary layer, where the vertical coordinate is rescaled as $\zeta=\delta^{-1} z=Ra^{1/10}z$, and the horizontal averages $\ov{\omega}$, $\ov{\psi}$, and $\ov{\theta}$ over a single roll are transformed into the $O(1)$ fields $\ov{\Omega^o}$, $\ov{\Psi^o}$, and $\ov{\Theta^o}$.  The collapse of these curves provides strong evidence for the existence of an outer boundary layer. Additionally, the top row is consistent with the predictions for the core region, since $\overline{\omega}$ and $\overline{\psi}$ become $z$-independent with increasing $Ra^{1/10}z$, and $\partial_z\ov{\theta}$ approaches $-K^4$.  Further analysis of the core region is unnecessary since \cref{fig:Contour_TPsiOmega,fig:Spectra_TPsiOmega} already demonstrate its heat-exchanger flow structure~\cref{eq:solution_C}.

The bottom row of \cref{fig:Profiles_variousBLs} demonstrates the collapse of the inner boundary layer at various $Ra$ values after rescaling according to \cref{tab:Order}.  The temperature profile appears nearly linear, consistent with the assumption made in the asymptotic analysis to conclude that $Nu=O(\eps^{-1})$.  By definition, the inner layer is a thermal boundary layer, as it is the only region where conduction contributes to the vertical heat flux at leading order.  However, it also contains a velocity boundary layer.  The stress-free boundary conditions impose that $\omega$ must vanish as $\eta = \eps^{-1}z = Ra^{3/10} z\to 0^+$, whereas no-slip conditions would instead require $\partial_z\psi$ to vanish.  Consequently, the velocity field would differ significantly in the inner boundary layer under no-slip conditions, although it might remain similar in other regions.

\begin{figure}[t]
\centering
\includegraphics[width=0.98\textwidth]{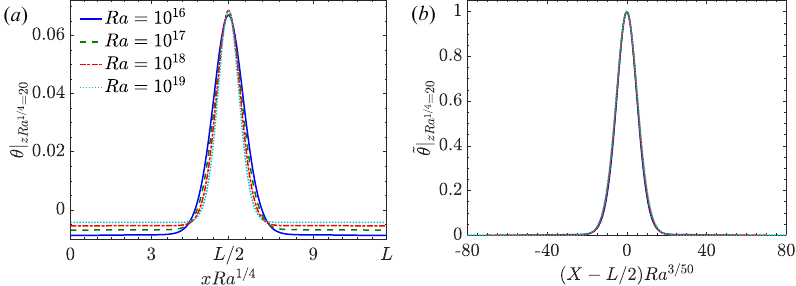}
\caption{Horizontal variation of ($a$) $\theta$ and ($b$) $\tilde\theta$ at rescaled height $zRa^{1/4}=20$ for different $Ra$ with $Pr=0.1$ and $k/Ra^{1/4}=0.5361$. The plotted quantity $\tilde\theta$ is a rescaling and translation of $\theta$, different at each $Ra$, so that its plotted values range from 0 to 1.  The collapse of $\tilde\theta$ profiles in ($b$) as a function of $\chi = (X-L/2)Ra^{3/50}$ for various $Ra$ strongly supports the predicted $O(Ra^{-3/50})$ scaling for the plume's width.}
\label{fig:theta_plumewidth}
\end{figure}

The middle row of \cref{fig:Profiles_variousBLs} highlights the collapse with varying (large) $Ra$ values in the middle boundary layer, where the vertical coordinate is defined as $Z=Ra^{1/4}z$.  The fields in panels ($d$) and ($e$) are scaled according to the bulk region of the middle layer, reflecting the fact that the bulk region occupies the vast majority of the layer's horizontal extent. The curves exhibit strong collapse across all three $Ra$, with minor imperfections likely due to the plume. For $10^{17}\le Ra \le 10^{19}$, the plume still occupies roughly one quarter of the roll width in the middle layer.  Within this range, the profiles of $\ov\psi$ collapse remarkably well when rescaled using an intermediate (though not fully asymptotic) scaling of $Ra^{-0.41}$, which lies between the predicted scalings for the plume region ($Ra^{-11/25}$) and the bulk ($Ra^{-2/5}$).  The influence of the plume diminishes gradually as $Ra$ increases---for example, it is expected to occupy less than one tenth of the roll width at $Ra = 10^{26}$, since its width, scaling as $O(Ra^{-3/50})$, narrows only slowly. Within the plume region, \cref{fig:Profiles_variousBLs}(\emph{$f$}) demonstrates the collapse of temperature profiles along the plume's centerline (located at the roll boundary $x=\Gamma/2$), consistent with the predicted scaling $\theta =O(Ra^{-1/100})$.  The predicted $O(Ra^{-3/50})$ scaling for the plume's width is further validated by \cref{fig:theta_plumewidth}, which shows how $\theta$ and its rescaled and translated form $\tilde\theta$ vary horizontally across the plume at a fixed rescaled height $zRa^{1/4}=20$.  At this rescaled height---and others---the horizontal temperature variation across the plume closely follows a Gaussian distribution. For instance, the collapsed curves in \cref{fig:theta_plumewidth}($b$) fit the Gaussian function $0.9851 e^{-(\tilde{x}/7.944)^2}$ with an R-squared value of $0.9996$, where $\tilde{x}$ denotes the coordinate on the horizontal axis in \cref{fig:theta_plumewidth}($b$).

\section{Regimes and transitions in the $Ra$--$k$ plane}
\label{sec:fixed Gamma}

The asymptotic structure of steady convection rolls can vary depending upon how $k$ and $Pr$ scale with $Ra$ as $Ra\to\infty$. With $Pr$ fixed, we have detailed one asymptotic structure in \cref{sec:LocaMax_numerics,sec:LocaMax_analysis} for $k = O(Ra^{1/4})$, whereas a distinct structure has been described for $k = O(1)$ in previous work \citep{ChiniCox2009, Wen2020JFM}.  While solutions featuring a heat-exchanger core~\cref{eq:solution_C} can only occur when $k = O(Ra^{1/4})$, the asymptotic structure presented by \cite{ChiniCox2009} arises for $k = O(Ra^{\beta_k})$ across a range of exponents.  In \cref{sec:regimes}, we identify regimes in the $Ra$--$k$ plane where these different asymptotic structures are found, focusing on $Pr = 1$ and $k \ge O(1)$.  In \cref{sec:transitions}, we illustrate transitions between these regimes as $Ra$ increases at fixed $k$.

\subsection{Parameter regimes}
\label{sec:regimes}

\begin{figure}[t]
\centering
\includegraphics[width=0.55\textwidth]{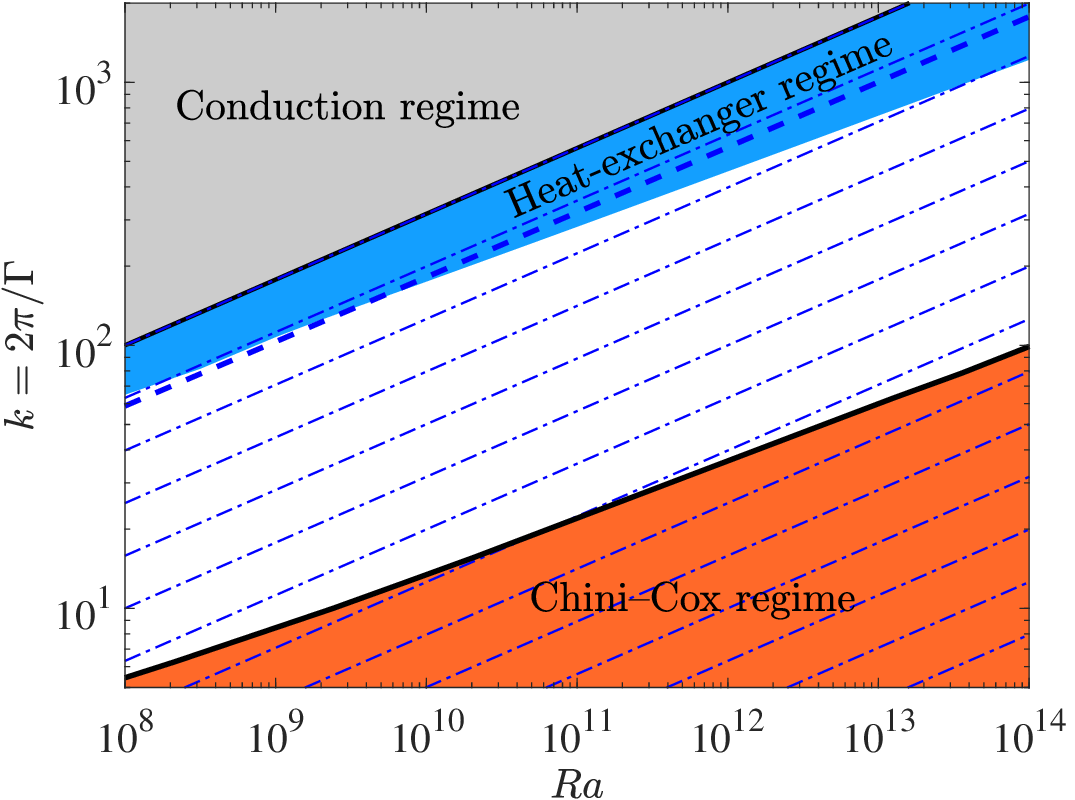}
\caption{Partial regime diagram of steady rolls for $k\ge O(1)$ at large $Ra$ with $Pr=1$ and stress-free boundaries. The heat-exchanger and Chini--Cox regimes are defined in the text.  At large $k$, the lower boundary of the conduction region scales as $k \sim Ra^{1/4}$ because the critical Rayleigh number satisfies $Ra_c \sim k^4$.  Dash-dotted lines (\blue{$\dashdottedrule$}) denote $k= c\,Ra^{1/4}$ for various $c$. The dashed line (\blue{$\dashedrule$}) denotes the locally $Nu$-maximizing wavenumbers $k^*_{loc}\approx 0.56\,Ra^{1/4}$. The upper boundary of the Chini--Cox regime is defined in the text; the lower boundary of the heat-exchanger regime remains to be determined precisely.}
\label{fig:kRa_solns}
\end{figure}

\Cref{fig:kRa_solns} shows a partial regime diagram in the $Ra$--$k$ plane for the large-$Ra$ structure of rolls with $Pr=1$.  In the ``conduction regime'', the conduction state is globally stable, so there are no roll states.  The ``heat-exchanger regime'' is the region of parameter space where the core structure of the rolls closely matches the heat-exchanger flow given by \cref{eq:heatexchanger}---a key component of the asymptotic framework proposed in \cref{sec:LocaMax_analysis}.  The ``Chini--Cox regime'' denotes the parameter regime where the rolls exhibit the asymptotic structure proposed by \cite{ChiniCox2009}: a dynamically inviscid core characterized by uniform temperature and vorticity, surrounded by narrow, decoupled thermal and viscous boundary layers and prominent upwelling and downwelling plumes along the cell perimeter. In between the heat-exchanger and Chini--Cox regimes, the rolls do not assume any known asymptotic structure.

The heat-exchanger regime in \cref{fig:kRa_solns} is bounded above by the solid line where $Ra = Ra_c$, whose exact expression~\cref{eq:Ra_c} gives $Ra \sim k^4$ at large $k$ or, equivalently, $k \sim Ra^{1/4}$. The shading of this regime is where the mean temperature gradient at the midplane $z = 1/2$ is within $1\%$ of its heat-exchanger value, $\partial_z\overline{T} = -k^4/Ra$. The lower boundary of this regime is difficult to define precisely at each $Ra$ because there can be several $k$ values at which the midplane temperature gradient differs from the heat-exchanger value by $1\%$. The boundary in the figure denotes the largest $k$ values at which this relative difference is $1\%$. In limits where $k \sim c\,Ra^{1/4}$---illustrated by dash-dotted lines in \cref{fig:kRa_solns}---we expect rolls to approach the asymptotic structure described in \cref{sec:LocaMax_analysis} for any fixed $c \in (0,1)$. This includes the locally $Nu$-maximizing solutions reported in \cref{sec:LocaMax_numerics}, which have $c \approx 0.56$ and are indicated by a dashed line in \cref{fig:kRa_solns}. When $c$ is closer to zero, larger $Ra$ are needed to reach the asymptotic heat-exchanger structure, which explains the downward curvature of the regime's approximate lower boundary. When $c = 1$ but $Ra > k^4$, meaning $Ra = k^4 + o(k^4)$, there is a sub-regime of the heat-exchanger regime that abuts the conduction regime and has rolls with simpler structure: only a heat-exchanger core and a single isotropic boundary layer. These rolls were studied using matched asymptotics by \citet{Blennerhassett1994}, who describe the solutions as strongly nonlinear. However, as explained in \cref{app:BB}, their asymptotics apply only to rolls sufficiently weak that the temperature field in the core approaches the conduction state as $k \to \infty$ with $Ra \sim Ra_c$.

The Chini--Cox regime in \cref{fig:kRa_solns} is where $Nu/Ra^{1/3}$ is within 2\% of the $k$-dependent value predicted by \citet{ChiniCox2009} for the $k=O(1)$ asymptotic structure summarized in the first paragraph of this subsection. Using 1\% as the criterion rather than 2\% would give an upper boundary with similar slope but that is much harder to compute at each $k$ because $Ra$ increases by an order of magnitude.  The upper boundary shown in \cref{fig:kRa_solns} scales approximately as $k\approx0.082\,Ra^{0.22}$. To delineate this boundary, we have extended the asymptotic predictions for the ratios $Nu/Ra^{1/3}$ and $Re/Ra^{2/3}$ from \citet{ChiniCox2009} to larger $k$, as detailed at the end of \cref{sec:transitions}. Rolls are classified as belonging to the Chini--Cox regime when both ratios fall within 2\% of their asymptotically predicted values; because the $Nu$ criterion is more restrictive, it gives the upper regime boundary in the figure. We have not identified the lower boundary of this regime since we have computed rolls only with $k \ge 0.5$. Finally, while the explicit $Pr$-dependence is known for both the Chini--Cox rolls and the heat-exchanger flow, how $Pr$ influences the boundaries of the Chini--Cox regime---and the lower boundary of the heat-exchanger regime---remains a subject for future investigation.

\subsection{Regime transitions with increasing $Ra$}
\label{sec:transitions}

\begin{figure}[p]
\centering
\includegraphics[width=0.84\textwidth]{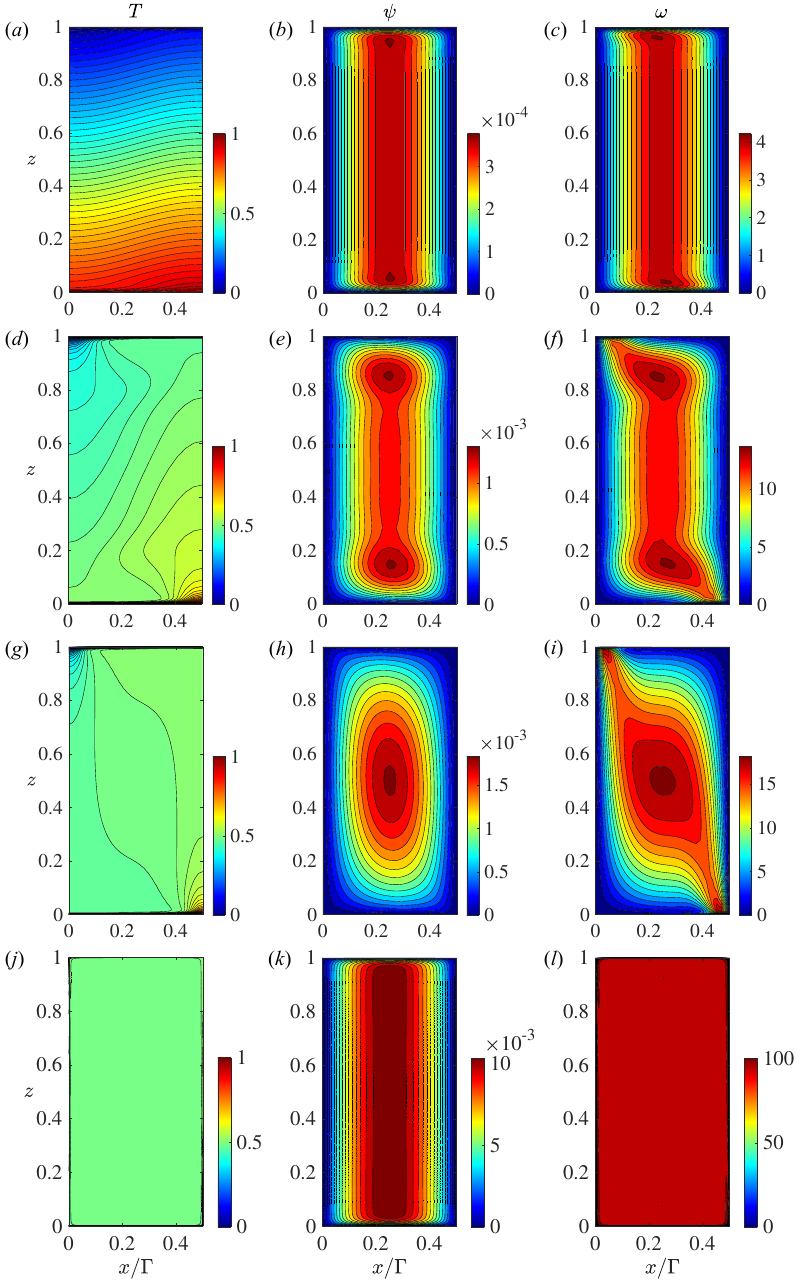}
\caption{Contours of $T$ (left), $\psi$ (middle), and $\omega$ (right) for steady rolls with $\Gamma=0.06$, $Pr=1$, and (from top to bottom) $Ra = 1.45\times10^8$, $10^{9}$, $10^{10}$, and $10^{14}$. 
Rolls of period $\Gamma=0.06$ (fundamental wavenumber $k\approx 105$) exist when $Ra>Ra_c\approx1.2 \times 10^8$.  Specifically, the solutions approximately correspond to: the Blennerhassett--Bassom regime (row 1); a state close to the locally $Nu$-maximizing period where $\Gamma^*_{\mathrm{loc}} \approx 0.061$ (row 2); a transition regime between the heat-exchanger and Chini--Cox structures (row 3); and the Chini--Cox regime (row 4).}
\label{fig:Contour_TPsiOmega_Gamma006}
\end{figure}

\Cref{fig:Contour_TPsiOmega_Gamma006} illustrates steady rolls with a fixed horizontal period of $\Gamma = 0.06$ across four Rayleigh numbers, increasing from $1.45 \times 10^8$ in the top row to $10^{14}$ in the bottom row.  Rolls with this period possess a fundamental horizontal wavenumber of $k = 2\pi/\Gamma \approx 105$ and bifurcate from the purely conduction state once $Ra$ exceeds the critical threshold $Ra_c \approx 1.2 \times 10^8$.  At the lowest Rayleigh number considered ($Ra = 1.45 \times 10^8$, top row), the roll solution lies slightly beyond the small-amplitude regime described by \citet{Blennerhassett1994}. At $Ra = 10^9$ (second row)---situated just past the lower boundary of the heat-exchanger regime ($Ra \approx 8.9 \times 10^8$ for $k \approx 105$)---the recirculation zones expand toward the core, even as the roll continues to display the three nested boundary layers detailed in \cref{sec:LocaMax_analysis}. This solution is also very close to a locally $Nu$-maximizing state ($\Gamma^*_{\mathrm{loc}} \approx 0.061$ at $Ra = 10^9$), which itself lies just outside the heat-exchanger regime.  At higher $Ra$, however, these locally $Nu$-maximizing rolls merge into the heat-exchanger regime and fully adopt the asymptotic structure described in \cref{sec:LocaMax_analysis}. As $Ra$ increases further, transitioning from the heat-exchanger regime toward the Chini--Cox regime, the rolls deform continuously from one asymptotic structure to the other. An intermediate state along this trajectory ($Ra = 10^{10}$) is captured in the third row of \cref{fig:Contour_TPsiOmega_Gamma006}. Finally, at the largest value investigated ($Ra = 10^{14}$, bottom row), the roll develops a nearly uniform temperature and vorticity core. This state sits just outside the Chini--Cox regime, which is estimated to begin at $Ra \gtrsim 1.3 \times 10^{14}$ for $k \approx 105$.

\begin{figure}[t!]
\centering
\includegraphics[width=0.98\textwidth]{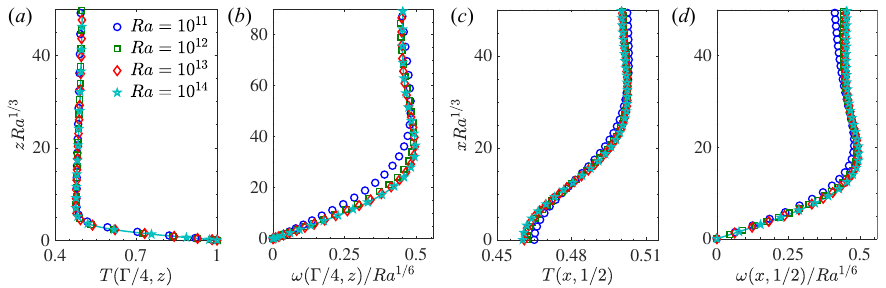}
\caption{Rescaled spatial structure of temperature $T$ and vorticity $\omega$ near the bottom $(a,b)$ and left $(c,d)$ boundaries of a counterclockwise roll with $Pr=1$ and $\Gamma=0.06$.  Solid curves are spectral interpolants of $10^{14}$ values.  For comparison, analogous plots at the larger aspect ratio $\Gamma = 2$ are shown in figure~5 of \citet{Wen2020JFM}, where a different scaling for $\omega$ was used due to the choice of diffusion velocity scale for nondimensionalization in that study.
}
\label{fig:MeanBL_TOmega_Gamma006}
\end{figure}
\begin{figure}[h!]
\centering
\includegraphics[width=0.98\textwidth]{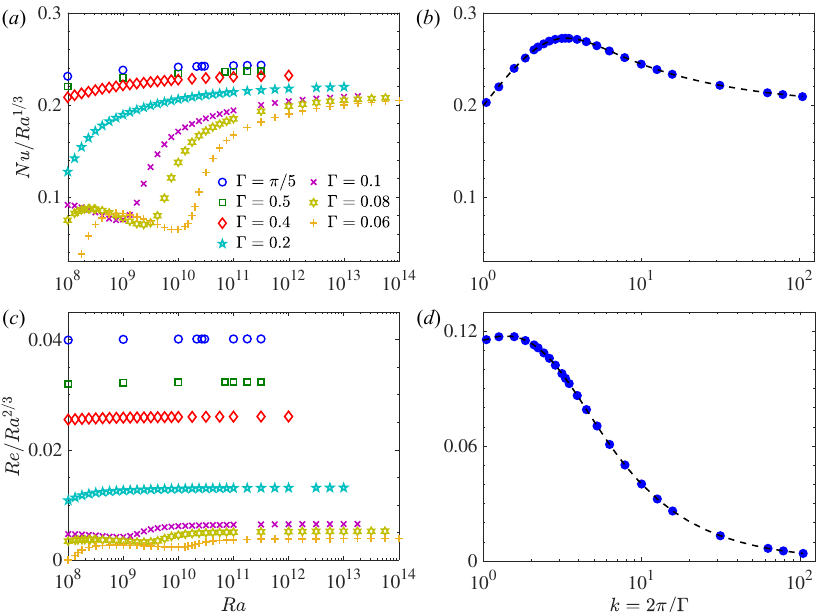}
 \caption{($a,c$) Dependence of the ratios $Nu/Ra^{1/3}$ and $Re/Ra^{2/3}$ on $Ra$ for various $\Gamma$ with $Pr=1$, and \mbox{($b,d$)} large-$Ra$ asymptotic values of the same ratios for various $k=2\pi/\Gamma$, estimated from numerically computed rolls (\textcolor{blue}{$\bullet$}) and from the asymptotic analysis of \citeauthor{ChiniCox2009} ($\dashedrule$). In panels ($b$) and ($d$), the bullet-point symbols for $k \geq 10$ denote the estimated asymptotes of the data series in panels ($a$) and ($c$), respectively, while those for $k < 10$ correspond to analogous estimates based on the data previously reported in figure~3 of \cite{Wen2020JFM}. The dashed lines in panels ($b$) and ($d$) represent the prefactors in the asymptotic relations $Nu\sim c_n(k)Ra^{1/3}$ and $Re\sim c_r(k)Pr^{-1}Ra^{2/3}$, respectively, as proposed by \citet{ChiniCox2009}.  Values of $c_n(k)$ and $c_r(k)$ were computed using the code from \cite{ChiniCox2009}.}
\label{fig:NuRe_Gammafixed}
\end{figure}

The asymptotic structure proposed by \citet{ChiniCox2009} features boundary layers on all four sides of the roll, whose thicknesses scale as $Ra^{-1/3}$ when $Ra$ increases at fixed $\Gamma$. \Cref{fig:MeanBL_TOmega_Gamma006} shows temperature and vorticity profiles near the boundaries at various $Ra$, with spatial coordinates rescaled by $Ra^{1/3}$. These rescaled profiles converge as $Ra$ increases from $10^{11}$ to $10^{14}$, consistent with the asymptotic predictions. Analogous plots for rolls with the larger aspect ratio $\Gamma=2$ are shown in figure 5 of \citet{Wen2020JFM}. At that larger $\Gamma$, the rescaled boundary layer fields are already close to their asymptotic forms when $Ra$ exceeds $10^6$. For the much smaller $\Gamma=0.06$ in \cref{fig:MeanBL_TOmega_Gamma006}, comparable convergence requires $Ra \gtrsim 10^{13}$.  We expect such convergence to occur for any fixed $\Gamma$, though smaller $\Gamma \le O(1)$---i.e., larger $k$---will require higher $Ra$ to reach the Chini--Cox regime.

\Cref{fig:NuRe_Gammafixed} illustrates how $Nu$ and $Re$ approach the asymptotic scalings predicted by \citeauthor{ChiniCox2009} as $Ra$ increases at fixed $\Gamma$.  The convergence of each data series toward horizontal lines in panels ($a$) and ($c$) aligns with the asymptotic predictions $Nu \sim c_n(k)Ra^{1/3}$ and $Re \sim c_r(k)Pr^{-1}Ra^{2/3}$, respectively.  Pre-asymptotic behaviour is evident at smaller $Ra$, with smaller $\Gamma$---equivalently larger $k$---requiring higher $Ra$ to reach the asymptotic regime, roughly $Ra \gtrsim 8.6 \times 10^4\, k^{0.46}$ (cf. \cref{fig:kRa_solns}).  Panels ($b$) and ($d$) display the asymptotic prefactors $c_n$ and $c_r$, respectively, as estimated from numerically computed rolls (bullet-point symbols) and as computed using the code of \cite{ChiniCox2009} (dashed lines).  The criterion used to define the Chini--Cox regime in the $Pr=1$ phase diagram of \cref{fig:kRa_solns} is that numerically computed rolls have both $Nu/Ra^{1/3}$ and $Re/Ra^{2/3}$ within 2\% of the corresponding dashed-line values.

\section{Conclusions}
\label{sec:conclusions}

Although convection at sufficiently large $Ra$ is invariably turbulent, coherent states can capture aspects of turbulent heat transport and provide insight into flow self-organisation. Recent investigations of 2D RBC between no-slip boundaries have revealed that steady roll solutions achieve greater heat transport at certain high wavenumbers \citep{Waleffe2015,Sondak2015,Wen2022JFM}. Here we have performed a systematic numerical and asymptotic study of the complementary stress-free case, focusing on small aspect ratio $\Gamma$---i.e., large wavenumber $k=2\pi/\Gamma$. For rolls computed with $10^8\le Ra \le 10^{19}$ and $10^{-1}\le Pr\le 10^{3/2}$, a locally $Nu$-maximizing solution emerges with $k \propto Ra^{1/4}$, exhibiting approximate scalings $Nu \propto Ra^{0.29}$ and $Re\,Pr \propto Ra^{0.40}$. These scaling exponents are essentially the same as in the no-slip case when $k \propto Ra^{1/4}$ with $Pr = 1$ \citep{Wen2022JFM}. These rolls develop a heat-exchanger core and near-wall recirculation zones of vertical thickness $O(Ra^{-0.1})$. In the stress-free case, the growth of $Nu$ with $Ra$ is slower when $k \propto Ra^{1/4}$ than when $k$ is fixed, which instead gives $Nu = O(Ra^{1/3})$.

As in a closely related analysis of steady axisymmetric Taylor vortices in Taylor--Couette flow \citep{Deguchi2023}, our matched asymptotic analysis of steady rolls with $k = O(Ra^{1/4})$ reveals a vertically stacked four-layer structure near each boundary. These rolls have a heat-exchanger core structure characterised by a single marginally-stable nonzero Fourier mode in the horizontal ($x$) direction---much like other steady quasilinear (sub)systems \citep{Herring1963,Michel2019,Pausch2019,OConnor2021,Wen2022PTRSA,Chini2022}---and a horizontal mean temperature profile that varies linearly in the vertical ($z$) direction. Since the $x$-varying core solution is $z$-independent, and the boundary conditions require no flow in the $z$ direction, there must be an isotropic inertial turnaround region, meaning it has negligible diffusion and commensurate $O(Ra^{-1/4})$ width and thickness---this is the middle boundary layer. The vertical velocity grows algebraically as $z^{1/3}$ in the far-field of this middle layer, so it cannot be matched directly with the core flow. Instead, an outer boundary layer of thickness $O(Ra^{-1/10})$ is required to connect the two. Finally, an inner thermal boundary layer of thickness $O(Ra^{-3/10})$, in which advection and vertical heat diffusion balance, accommodates the fixed temperature condition at each wall. Since most of the $O(1)$ temperature drop across the domain occurs in these inner layers, the $Ra$-scaling of $Nu$ is set by their inverse thickness, yielding $Nu = O(Ra^{3/10})$.

Our asymptotic analysis differs from the work of \cite{Deguchi2023} in several ways. Most importantly, we find that a thin vertical plume is required along one side of the middle boundary layer for successful matching of the heat flux between the inner and outer boundary layers. Outside of this plume, the temperature in the middle boundary layer is constant to leading order, implying negligible heat flux across this subdomain. This constant temperature is consistent with the downwelling flow in the outer boundary layer satisfying laminar plume equations in a rescaled domain that is semi-infinite in the vertical `streamwise' direction. In particular, the thermal anomaly advected from the core to the outer boundary layer can diffuse over an infinite distance (in rescaled variables) before reaching the lower wall. In the middle layer, however, it is the plume that carries the majority of the heat flux. The inclusion of this plume region is consequential not only for the structure of high-wavenumber rolls but also for the scaling of the resulting heat transport. Unlike \citet{Deguchi2023}, our predicted scalings do not include logarithmic modifications. Furthermore, our analysis of the inner boundary layer finds buoyancy torque arising at leading order, along with advection and vertical diffusion of vorticity. This differs from \citet{Deguchi2023}, and from the asymptotics of rolls with $k=O(1)$ and stress-free boundaries \citep{ChiniCox2009}, where buoyancy torque is significant only in domain-spanning plumes, and where $Nu$ scales differently from the no-slip case. We speculate that buoyancy torque arising at leading order in the inner boundary layer in the $k = O(Ra^{1/4})$ limit is why the same $Nu = O(Ra^{3/10})$ scaling seems to occur with both stress-free and no-slip boundaries.

The asymptotic structure we have proposed extends the analysis of \citet{Blennerhassett1994} to strongly nonlinear states that are valid in a much broader parameter regime. At fixed large $k$, when $Ra$ slightly exceeds the conduction state's linear stability threshold $Ra_c \sim k^4$, the flow is weakly nonlinear in the conventional sense: the dominant $x$-varying thermal perturbation of size $a \ll 1$ is asymptotically larger than the $O(a^2)$ correction to the horizontally averaged temperature. For somewhat larger $Ra$ satisfying $Ra = k^4 + o(k^4)$, \cite{Blennerhassett1994} identified rolls with a two-region vertical structure: a heat-exchanger core and one boundary layer of thickness $O(Ra^{-1/4})$ at each wall. This boundary layer is an isotropic, diffusive turnaround region in which advection, diffusion and buoyancy all contribute at leading order, so its rescaled equations are identical to the full equations. The solutions of \citeauthor{Blennerhassett1994} are more nonlinear than conventional weakly nonlinear states in the sense that the $x$-varying thermal perturbation and the correction to the horizontally averaged temperature are both $O(1/k)$ (cf.\ appendix~\ref{app:BB}). However, their solutions are still weakly nonlinear in the sense that the distortion of the $O(1)$ conduction profile is asymptotically small. The roll structure we have proposed, on the other hand, has the same $\Gamma=O(Ra^{-1/4})$ scaling but is strongly nonlinear.

Our numerical computations clarify the changes in flow structure between different regimes as $Ra$ is raised with small but fixed aspect ratio $\Gamma = 2\pi/k$. First, rolls quickly depart the asymptotically narrow regime of \citeauthor{Blennerhassett1994} and enter the strongly nonlinear heat-exchanger regime, within which rolls pass through a locally $Nu$-maximizing combination of $Ra$ and $k$. This transition is marked by a near-wall recirculation zone developing and moving away from the boundary, corresponding to the emergence of the outer boundary layer and the weakening of diffusive effects within the isotropic turnaround region. In this way, the single isotropic boundary layer of the \citeauthor{Blennerhassett1994} solution morphs into the middle boundary layer of our four-layer structure. Raising $Ra$ further causes the plume within the middle boundary layer to strengthen and expand, eventually obliterating the outer boundary layer. Concurrently, buoyancy torque weakens within the inner thermal boundary layer, and the core temperature and vorticity homogenise as advection dominates diffusion. The flow then ultimately transitions into the regime described by \citet{ChiniCox2009}, characterised by $Nu = O(Ra^{1/3})$ and $Re\,Pr = O(Ra^{2/3})$.

Although the present work illuminates the structure of convection rolls in one natural parameter limit, significant open questions remain about the asymptotic properties of convection rolls. One question is how the Prandtl number affects asymptotic scalings. Another question is what differs between stress-free and no-slip boundaries in the present $k = O(Ra^{1/4})$ limit, given that the scaling exponents appear to be the same. Finally, with no-slip boundaries, it remains to understand the asymptotic structure of rolls where \mbox{$k\to\infty$} at the rate that globally maximizes $Nu$. This limit is slower than $k= O(Ra^{1/4})$ but has been proposed to be both $O(Ra^{1/5})$ \citep{Wen2022JFM} and $O(Ra^{2/9})$ \citep{Deguchi2023}. Rolls in this $Nu$-maximizing limit are perhaps the most relevant to turbulent convection with no-slip boundaries, given their striking coincidence with turbulent $Nu$ values. To further explore whether the asymptotic scaling of rolls gives insight into turbulent heat transport, the asymptotic structure of rolls in this $Nu$-maximizing limit---and the proper scaling of their aspect ratio---must be elucidated.

\vspace{-0.in}
\section*{Acknowledgements}
We thank Kengo Deguchi for useful discussions. Some of this work was carried out at the Geophysical Fluid Dynamics Program of the Woods Hole Oceanographic Institution (NSF OCE-1829864). BW and GPC were supported by US National Science Foundation (award DMS-2532634). DG was supported by the NSERC Discovery Grants Program (awards RGPIN-2018-04263 and RGPIN-2025-06823).  Computational resources were provided by Advanced Research Computing at the University of Michigan and by the Digital Research Alliance of Canada.

\vspace{-0.in}
\section*{Declaration of interests}
The authors report no conflict of interest.
\vspace{-0.in}

\appendix
\crefalias{section}{appsec}
\crefalias{subsection}{appsec}
\crefalias{subsubsection}{appsec}

\section{\label{app:BB}Relation to asymptotics of \citeauthor{Blennerhassett1994}}

\cite{Blennerhassett1994} performed weakly nonlinear asymptotic analysis of convection rolls in the high-wavenumber limit of Rayleigh--B\'enard convection. Their asymptotic expansion is valid when $Ra$ is only slightly above the onset of rolls at $Ra_c$ in the sense that $Ra-Ra_c\ll Ra_c$. In particular, they expand as
\begin{equation}
\label{eq:Ra_BB}
Ra = k^4 + R_1k^3 + \cdots, \quad \text{as}\; k \to \infty,  
\end{equation}
where $R_1$ is any positive constant, and we recall that $Ra_c\sim k^4$ as $k\to\infty$. This expansion leads to an interior structure that is consistent with the heat-exchanger structure of \cref{sec:LocaMax_analysis}, but its range of validity in parameter space is much smaller.

In this appendix we use the nondimensionalization of \citet{Blennerhassett1994}, where the chosen time scale is the thermal diffusion scale $h^2/\kappa$ rather than the free-fall timescale we have used above. In thermal diffusion units we denote velocities as $\tilde{\mathbf{u}} = \tilde{u}\mathbf{\hat{x}} + \tilde{w}\mathbf{\hat{z}}$. Based on the expansion of the Rayleigh number in~\cref{eq:Ra_BB}, the asymptotic analysis conducted by \cite{Blennerhassett1994} yields the following expansions for the temperature and velocity fields in the core:
\begin{subequations}
\begin{align}
    T &=1 - z + k^{-1}\bar{\theta}_1 + k^{-1}[\theta_{10}(z)+k^{-1}\theta_{11}(z) +\cdots]\cos{kx} \nonumber\\
    &\quad + k^{-3}\theta_{20}(z)\cos{2kx}+\cdots, \\
    \tilde{u} &= [\tilde{u}_{10}(z) + k^{-1}\tilde{u}_{11}(z) + \cdots]\sin{kx}+k^{-2}\tilde{u}_{20}(z)\sin{2kx}+\cdots, \\
   \tilde{w} &= k[\tilde{w}_{10}(z) + k^{-1}\tilde{w}_{11}(z) + \cdots]\cos{kx}+k^{-1}\tilde{w}_{20}(z)\cos{2kx}+\cdots.
\end{align}
\end{subequations}
By substituting the known values of $\bar{\theta}_1 = R_1(z-1/2)$, $\theta_{10}=2\Upsilon$, $\tilde{u}_{10}=0$, and $\tilde{w}_{10}=2\Upsilon$ from \cite{Blennerhassett1994} and retaining only the leading-order terms in $k$ as $k \rightarrow \infty$, we obtain the following approximation for the temperature, stream function, and vorticity in terms of the diffusion velocity scale:
\begin{subequations}
\label{eq:BBsoln}
\begin{align}
    T &\sim1-z + k^{-1}R_1(z-1/2)+2\Upsilon k^{-1}\cos{kx}, \\
    \tilde{\psi} &\sim-2\Upsilon\sin{kx},\\
    \tilde{\omega} &\sim-2\Upsilon k^{2}\sin{kx}.
\end{align}
\end{subequations}
Rewriting the heat-exchanger solution~\cref{eq:heatexchanger} in diffusion units and then substituting the expansion of $Ra$ from~\cref{eq:Ra_BB} yields
\begin{subequations}
\label{eq:HeatExchangersoln_diffusionscale}
\begin{align}
    T &=1-z+k^{-1}R_1(z-1/2)\left(\frac{k}{k+R_1}\right)-\tilde{A}\sqrt{\frac{Pr}{k^4+R_1k^3}}k^3\cos{kx},\\
    \tilde{\psi} &=\tilde{A}\sqrt{Pr(k^4+R_1k^3)}\sin{kx},\\
    \tilde{\omega} &=\tilde{A}\sqrt{Pr(k^4+R_1k^3)}k^2\sin{kx}.
\end{align}
\end{subequations}
Choosing $\Upsilon=-{\tilde{A}\sqrt{Pr}k^2}/{2}$, we see that the heat-exchanger solution~\cref{eq:HeatExchangersoln_diffusionscale} and the Blennerhassett--Bassom solution~\cref{eq:BBsoln} are asymptotically identical as $k\to\infty$. However, the analysis of Blennerhassett--Bassom relies on the expansion~\cref{eq:Ra_BB} that is valid only for $Ra$ asymptotically close to $Ra_c$.

\section{\label{appB:Data}Numerical solutions}

\Cref{tab:GammaLocal} lists the values of $k^*_{\mathrm{loc}}$, $Nu$, and $Re$ obtained from numerical solutions across various combinations of $Pr$ and $Ra$. This subset of our results includes all data plotted in~\cref{fig:NuLocPrfixed}, along with values for $Pr = 10^{1/4}$, $10^{1/2}$, and $10^{3/4}$ that are omitted from the figure for clarity.

\begin{longtable}{cccccc}
\caption{Details for numerical solutions with the aspect ratios $\Gamma^*_{loc}$ that locally maximize $Nu(\Gamma)$, including the resolution of Fourier modes ($N_x$) and Chebyshev collocation points ($N_z$).}
\label{tab:GammaLocal}
\\\quad $Pr$ \quad &	\quad $Ra$ \quad &	\quad  $k^*_{loc}=2\pi/\Gamma^*_{loc}$ \quad &	\quad\quad $N_x \times N_z$	\quad\quad &	\quad\quad\quad $Nu$ \quad\quad\quad &	\quad\quad\quad		$Re$	\quad\quad\quad\\
$10^{-1}$	&	$10^{13}$		&	951.3635	&	128 $\times$ 449	&	938.5617	&	1016596.1	\\
$10^{-1}$	&	$10^{53/4}$	&	1099.940	&	128 $\times$ 513	&	1113.628	&	1277422.1	\\
$10^{-1}$	&	$10^{54/4}$	&	1270.948	&	128 $\times$ 513	&	1321.467	&	1606181.7	\\
$10^{-1}$	&	$10^{55/4}$	&	1468.103	&	128 $\times$ 513	&	1568.198	&	2020188.3	\\
$10^{-1}$	&	$10^{14}$		&	1695.409	&	128 $\times$ 513	&	1861.070	&	2541577.9	\\
$10^{-1}$	&	$10^{57/4}$	&	1957.867	&	128 $\times$ 641	&	2208.742	&	3197619.3	\\
$10^{-1}$	&	$10^{58/4}$	&	2260.871	&	128 $\times$ 641	&	2621.399	&	4023143.1	\\
$10^{-1}$	&	$10^{59/4}$	&	2610.597	&	128 $\times$ 641	&	3111.116	&	5062160.7	\\
$10^{-1}$	&	$10^{15}$		&	3014.819	&	128 $\times$ 769	&	3692.536	&	6368537.2	\\
$10^{-1}$	&	$10^{61/4}$	&	3481.568	&	128 $\times$ 769	&	4382.407	&	8012273.1	\\
$10^{-1}$	&	$10^{62/4}$	&	4020.209	&	128 $\times$ 769	&	5200.861	&	10081794		\\
$10^{-1}$	&	$10^{63/4}$	&	4643.754	&	192 $\times$ 1537	&	6173.213	&	12679422		\\
$10^{-1}$	&	$10^{16}$		&	5363.871	&	192 $\times$ 1793	&	7326.513	&	15947914		\\
1		&	$10^{8}$		&	58.80539	&	128 $\times$ 129	&	42.66446	&	1081.3023	\\
1		&	$10^{9}$		&	103.1969	&	128 $\times$ 257	&	81.95752	&	2725.9285	\\
1		&	$10^{10}$		&	181.6590	&	128 $\times$ 257	&	158.2679	&	6842.2250	\\
1		&	$10^{41/4}$	&	209.1934	&	128 $\times$ 321	&	186.6922	&	8613.8643	\\
1		&	$10^{42/4}$	&	241.0796	&	128 $\times$ 321	&	220.2694	&	10836.387	\\
1		&	$10^{43/4}$	&	277.8670	&	128 $\times$ 321	&	259.9377	&	13630.452	\\
1		&	$10^{11}$		&	320.3141	&	128 $\times$ 321	&	306.8066	&	17142.679	\\
1		&	$10^{45/4}$	&	369.2950	&	128 $\times$ 385	&	362.1876	&	21557.342	\\
1		&	$10^{46/4}$	&	425.8141	&	128 $\times$ 385	&	427.6320	&	27106.271	\\
1		&	$10^{47/4}$	&	491.0350	&	128 $\times$ 385	&	504.9745	&	34080.546	\\
1		&	$10^{12}$		&	566.3027	&	128 $\times$ 385	&	596.3847	&	42845.764	\\
1		&	$10^{49/4}$	&	653.1649	&	128 $\times$ 449	&	704.4287	&	53861.686	\\
1		&	$10^{50/4}$	&	753.4157	&	128 $\times$ 449	&	832.1421	&	67705.402	\\
1		&	$10^{51/4}$	&	869.1037	&	128 $\times$ 449	&	983.1146	&	85104.074	\\
1		&	$10^{13}$		&	1002.631	&	128 $\times$ 449	&	1161.594	&	106968.02	\\
1		&	$10^{53/4}$	&	1156.741	&	128 $\times$ 513	&	1372.603	&	134443.76	\\
1		&	$10^{54/4}$	&	1334.605	&	128 $\times$ 513	&	1622.089	&	168971.98	\\
1		&	$10^{55/4}$	&	1539.883	&	128 $\times$ 513	&	1917.091	&	212362.89	\\
1		&	$10^{14}$		&	1776.668	&	128 $\times$ 513	&	2265.834	&	266911.93		\\
1		&	$10^{57/4}$	&	2050.380	&	128 $\times$ 641	&	2678.354	&	335394.15	\\
1		&	$10^{58/4}$	&	2366.013	&	128 $\times$ 641	&	3166.133	&	421498.64	\\
1		&	$10^{59/4}$	&	2730.158	&	128 $\times$ 641	&	3742.773	&	529737.08	\\
1		&	$10^{15}$		&	3151.362	&	128 $\times$ 641	&	4424.663	&	665562.43	\\
1		&	$10^{61/4}$	&	3636.537	&	128 $\times$ 769	&	5231.827	&	836464.39	\\
1		&	$10^{62/4}$	&	4196.904	&	128 $\times$ 833	&	6186.067	&	1051136.7		\\
1		&	$10^{63/4}$	&	4844.324	&	192 $\times$ 1537	&	7314.996	&	1320714.9	\\
1		&	$10^{16}$		&	5589.824	&	192 $\times$ 1793	&	8650.078	&	1660004.0	\\
$10^{1/4}$	&	$10^{13}$ 	&	1014.399	&	128 $\times$ 449	&	1261.020	&	61942.286	\\
$10^{1/4}$	&	 $10^{53/4}$ 	&	1170.118	&	128 $\times$ 513	&	1488.384	&	77821.793	\\
$10^{1/4}$	&	 $10^{54/4}$ 	&	1349.829	&	128 $\times$ 513	&	1756.946	&	97768.891	\\
$10^{1/4}$	&	 $10^{55/4}$ 	&	1557.094	&	128 $\times$ 513	&	2074.184	&	122836.07	\\
$10^{1/4}$	&	 $10^{14}$ 	&	1796.120	&	128 $\times$ 513	&	2448.789	&	154341.53	\\
$10^{2/4}$	&	 $10^{13}$ 	&	1022.471	&	128 $\times$ 449	&	1408.850	&	36534.933	\\
$10^{2/4}$	&	 $10^{53/4}$ 	&	1179.742	&	128 $\times$ 513	&	1661.486	&	45869.716	\\
$10^{2/4}$	&	 $10^{54/4}$ 	&	1361.205	&	128 $\times$ 513	&	1959.638	&	57590.557	\\
$10^{2/4}$	&	 $10^{55/4}$ 	&	1570.404	&	128 $\times$ 513	&	2311.411	&	72316.238	\\
$10^{2/4}$	&	 $10^{14}$ 	&	1811.813	&	128 $\times$ 513	&	2726.340	&	90806.290	\\
$10^{3/4}$	&	 $10^{13}$ 	&	1023.870	&	128 $\times$ 449	&	1602.879	&	21889.052	\\
$10^{3/4}$	&	 $10^{53/4}$ 	&	1182.406	&	128 $\times$ 513	&	1889.953	&	27454.846	\\
$10^{3/4}$	&	 $10^{54/4}$ 	&	1365.198	&	128 $\times$ 513	&	2228.462	&	34442.334	\\
$10^{3/4}$	&	 $10^{55/4}$ 	&	1576.037	&	128 $\times$ 513	&	2627.391	&	43213.803	\\
$10^{3/4}$	&	 $10^{14}$ 	&	1820.263	&	128 $\times$ 513	&	3097.981	&	54199.438	\\
10			&	 $10^{13}$ 	&	1015.497	&	128 $\times$ 449	&	1832.490	&	13270.719	\\
10			&	 $10^{53/4}$ 	&	1174.581	&	128 $\times$ 513	&	2162.404	&	16625.476	\\
10			&	 $10^{54/4}$ 	&	1357.998	&	128 $\times$ 513	&	2551.077	&	20835.583	\\
10			&	 $10^{55/4}$ 	&	1570.286	&	128 $\times$ 513	&	3009.156	&	26105.622	\\
10			&	 $10^{14}$ 	&	1816.264	&	128 $\times$ 513	&	3550.987	&	32756.655	\\
10			&	 $10^{57/4}$ 	&	2097.471	&	128 $\times$ 705	&	4187.088	&	41012.011		\\
10			&	 $10^{58/4}$ 	&	2423.694	&	128 $\times$ 769	&	4938.081	&	51409.516	\\
10			&	 $10^{59/4}$ 	&	2800.243	&	128 $\times$ 833	&	5823.416	&	64449.685	\\
10			&	 $10^{15}$ 	&	3235.087	&	192 $\times$ 1025	&	6867.102	&	80799.095	\\
10			&	 $10^{61/4}$ 	&	3736.542	&	192 $\times$ 1281	&	8097.689	&	101321.12	\\
10			&	 $10^{62/4}$ 	&	4315.375	&	192 $\times$ 1281	&	9548.612	&	127055.85	\\
10			&	 $10^{63/4}$ 	&	4983.317	&	192 $\times$ 1537	&	11259.49	&	159347.14	\\
10			&	 $10^{16}$ 	&	5754.644	&	192 $\times$ 1793	&	13276.93	&	199843.74	\\
$10^{5/4}$	&	 $10^{13}$ 	&	994.0491	&	128 $\times$ 577	&	2081.951	&	8124.2667	\\
$10^{5/4}$	&	 $10^{53/4}$ 	&	1152.159	&	128 $\times$ 577	&	2461.657	&	10167.572	\\
$10^{5/4}$	&	 $10^{54/4}$ 	&	1334.917	&	128 $\times$ 577	&	2909.579	&	12727.005	\\
$10^{5/4}$	&	 $10^{55/4}$ 	&	1546.402	&	128 $\times$ 641	&	3437.971	&	15930.290	\\
$10^{5/4}$	&	 $10^{14}$ 	&	1790.591	&	128 $\times$ 641	&	4060.774	&	19946.051	\\
$10^{5/4}$	&	 $10^{57/4}$ 	&	2073.453	&	192 $\times$ 896	&	4795.814	&	24966.961	\\
$10^{5/4}$	&	 $10^{58/4}$ 	&	2399.994	&	192 $\times$ 896	&	5662.227	&	31261.545	\\
$10^{5/4}$	&	 $10^{59/4}$ 	&	2777.221	&	192 $\times$ 961	&	6683.653	&	39148.383	\\
$10^{5/4}$	&	 $10^{15}$ 	&	3212.919	&	192 $\times$ 1281	&	7887.670	&	49031.644	\\
$10^{5/4}$	&	 $10^{61/4}$ 	&	3716.102	&	192 $\times$ 1281	&	9307.022	&	61418.068	\\
$10^{5/4}$	&	 $10^{62/4}$ 	&	4297.370	&	192 $\times$ 1409	&	10980.11	&	76939.387	\\
$10^{5/4}$	&	 $10^{63/4}$ 	&	4968.516	&	192 $\times$ 1501	&	12952.20	&	96395.692	\\
$10^{5/4}$	&	 $10^{16}$ 	&	5743.629	&	192 $\times$ 1793	&	15276.78	&	120781.57	\\
$10^{6/4}$	&	 $10^{13}$ 	&	957.8318	&	128 $\times$ 641	&	2328.866	&	5012.0065	\\
$10^{6/4}$	&	 $10^{53/4}$ 	&	1112.718	&	128 $\times$ 641	&	2762.322	&	6268.6104	\\
$10^{6/4}$	&	 $10^{54/4}$ 	&	1292.305	&	128 $\times$ 641	&	3274.915	&	7840.1147		\\
$10^{6/4}$	&	 $10^{55/4}$ 	&	1500.283	&	128 $\times$ 641	&	3880.552	&	9806.9930	\\
$10^{6/4}$	&	 $10^{14}$ 	&	1741.508	&	128 $\times$ 641	&	4595.677	&	12265.077	\\
$10^{6/4}$	&	 $10^{57/4}$ 	&	2021.032	&	192 $\times$ 896	&	5441.686	&	15339.430	\\
$10^{6/4}$	&	 $10^{58/4}$ 	&	2344.472	&	192 $\times$ 1024	&	6439.707	&	19187.559	\\
$10^{6/4}$	&	 $10^{59/4}$ 	&	2718.817	&	192 $\times$ 1281	&	7617.743	&	24003.207	\\
$10^{6/4}$	&	 $10^{15}$ 	&	3151.994	&	192 $\times$ 1281	&	9007.807	&	30030.034	\\
$10^{6/4}$	&	 $10^{61/4}$ 	&	3653.227	&	192 $\times$ 1281	&	10647.73	&	37572.563	\\
$10^{6/4}$	&	 $10^{62/4}$ 	&	4232.811	&	192 $\times$ 1537	&	12582.00	&	47015.779	\\
$10^{6/4}$	&	 $10^{63/4}$ 	&	4902.993	&	192 $\times$ 1537	&	14863.02	&	58838.671	\\
$10^{6/4}$	&	 $10^{16}$ 	&	5676.893	&	192 $\times$ 1793	&	17552.59	&	73653.921	
\end{longtable}

\bibliographystyle{jfm}
\bibliography{RBC}

@article{Eckhardt2020,
  title={Exact relations between {Rayleigh--B{\'e}nard} and rotating plane {Couette} flow in two dimensions},
  author={Eckhardt, B. and Doering, C. R. and Whitehead, J. P.},
  journal={J. Fluid Mech.},
  volume={903},
  pages={R4},
  year={2020},
}

@article{Herring1963,
  title={Investigation of problems in thermal convection},
  author={Herring, J. R.},
  journal={J. Atmos. Sci.},
  volume={20},
  pages={325--338},
  year={1963}
}

@article{Michel2019,
  title={Multiple scales analysis of slow--fast quasi-linear systems},
  author={Michel, G. and Chini, G. P.},
  journal={Proc. R. Soc. A},
  volume={475},
  pages={20180630},
  year={2019}
}

@article{Pausch2019,
  title={Quasilinear approximation for exact coherent states in parallel shear flows},
  author={Pausch, M. and Yang, Q. and Hwang, Y. and Eckhardt, B.},
  journal={Fluid Dyn. Res.},
  volume={51},
  pages={011402},
  year={2019},
}

@article{Chini2022,
  title={Exploiting self-organized criticality in strongly stratified turbulence},
  author={Chini, G. P. and Michel, G. and Julien, K. and Rocha, C. B. and Caulfield, C. P.},
  journal={J. Fluid Mech.},
  volume={933},
  pages={A22},
  year={2022},
  publisher={Cambridge University Press}
}

@article{Wen2022PTRSA,
  title={{Heat transport in Rayleigh--B{\'e}nard convection with linear marginality}},
  author={Wen, B. and Ding, Z. and Chini, G. P. and Kerswell, R. R.},
  journal={Phil. Trans. R. Soc. A},
  volume={380},
  pages={20210039},
  year={2022},
}

@article{OConnor2021,
  title={{Marginally stable thermal equilibria of Rayleigh--B{\'e}nard convection}},
  author={O'Connor, L. and Lecoanet, D. and Anders, E. H.},
  journal={Phys. Rev. Fluids},
  volume={6},
  pages={093501},
  year={2021},
}

@article{Feng2025,
  title={Steady solutions of {Rayleigh--B\'enard} convection between {N}avier-slip boundaries},
  author={Feng, Z.},
  journal={Phys. Rev. Fluids},
  volume={10},
  pages={053502},
  year={2025},
}

@ARTICLE{Rayleigh1916,
  author = {Rayleigh, Lord},
  title = {On convection currents in a horizontal layer of fluid, when the higher temperature is on the under side},
  journal = {Philos. Mag.},
  volume = {32},
  pages = {529--546},
  year = {1916}
}

@ARTICLE{Kadanoff2001,
  author = {Kadanoff, L. P.},
  title = {Turbulent Heat Flow: Structures and Scaling},
  journal = {Phys. Today},
  volume = {54},
  pages = {34--39},
  year = {2001}
}

@article{Doering2020,
author = {Doering, C. R.},
journal = {Proc. Natl. Acad. Sci. USA},
pages = {9671--9673},
title = {{Turning up the heat in turbulent thermal convection}},
volume = {117},
year = {2020}
}

@article{ChiniCox2009,
author = {Chini, G. P.  and Cox, S. M.},
title = {{Large Rayleigh number thermal convection: Heat flux predictions and strongly nonlinear solutions}},
journal = {Phys. Fluids},
volume = {21},
pages = {083603},
year = {2009},
}

@article{Waleffe2015,
author = {Waleffe, F. and Boonkasame, A. and Smith, L. M.},
journal = {Phys. Fluids},
pages = {051702},
title = {{Heat transport by coherent Rayleigh--B{\'{e}}nard convection}},
volume = {27},
year = {2015}
}

@article{Sondak2015,
author = {Sondak, D. and Smith, L. M. and Waleffe, F.},
journal = {J. Fluid Mech.},
pages = {565--595},
title = {{Optimal heat transport solutions for Rayleigh--B{\'{e}}nard convection}},
volume = {784},
year = {2015}
}

@article{Ding2021JFM,
  title={Coherent heat transport in two-dimensional penetrative {Rayleigh--B\'enard} convection},
  author={Ding, Z. and Wu, J.},
  journal={J. Fluid Mech.},
  volume={920},
  pages={A48},
  year={2021},
}

@article{Kooloth2021,
  title = {{Coherent solutions and transition to turbulence in two-dimensional Rayleigh--B\'enard convection}},
  author = {Kooloth, P. and Sondak, D. and Smith, L. M.},
  journal = {Phys. Rev. Fluids},
  volume = {6},
  pages = {013501},
  year = {2021}
}

@article{Motoki2021, 
title={{Multi-scale steady solution for Rayleigh--B{\'e}nard convection}}, 
author={Motoki, S. and Kawahara, G. and Shimizu, M.}, 
journal={J. Fluid Mech.}, 
volume={914},
pages={A14}, 
year={2021}, 
publisher={Cambridge University Press}
}

@article{Motoki2022,
  title={Steady thermal convection representing the ultimate scaling},
  author={Motoki, S. and Kawahara, G. and Shimizu, M.},
  journal={Phil. Trans. R. Soc. A},
  volume={380},
  pages={20210037},
  year={2022},
  publisher={The Royal Society}
}

@ARTICLE{Whitehead2011,
  author = {Whitehead, J. P. and Doering, C. R.},
  title = {Ultimate state of two-dimensional {Rayleigh--B\'{e}nard} convection between free-slip fixed-temperature boundaries},
  journal = {Phys. Rev. Lett.},
  year = {2011},
  volume = {106},
  pages = {244501}
}

@article{Wen2020JFM,
author = {Wen, B. and Goluskin, D. and LeDuc, M. and Chini, G. P. and Doering, C. R.},
journal = {J. Fluid Mech.},
pages = {R4},
volume = {905},
title = {{Steady Rayleigh--B{\'{e}}nard convection between stress-free boundaries}},
year = {2020}
}

@article{Wen2022JFM,
author = {Wen, B. and Goluskin, D. and Doering, C. R.},
journal = {J. Fluid Mech.},
pages = {R4},
volume = {933},
title = {{Steady Rayleigh--B{\'{e}}nard convection between no-slip boundaries}},
year = {2022}
}

@article{Ouyang2025,
  title={Touching the classical scaling in penetrative convection},
  author={Ouyang, Z. and Wang, Q. and Li, K. and Wen, B. and Ding, Z.},
  journal={Proc. Natl. Acad. Sci. USA},
  volume={122},
  pages={e2418468122},
  year={2025},
}

@Article{Wen2015JFM,
  author = {B. Wen and L. T. Corson and G. P. Chini},
  title = {Structure and stability of steady porous medium convection at large {R}ayleigh number},
  journal =  {J. Fluid Mech.},
  year =  {2015},
  volume =  {772},
  pages =  {197--224},
}

@Article{WenChini2018JFM,
  author = {B. Wen and G. P. Chini},
  title = {{Inclined porous medium convection at large Rayleigh number}},
  journal =  {J. Fluid Mech.},
  year =  {2018},
  volume = {837}, 
  pages = {670--702},
  }

@article{Deguchi2023,
  title={{On high-Taylor-number Taylor vortices}},
  author={Deguchi, K.},
  journal={J. Fluid Mech.},
  volume={967},
  pages={A11},
  year={2023}
}

@article{He2026,
  title={High-{R}ayleigh-number asymptotic classical scaling in three-dimensional steady natural convection},
  author={He, X. and Motoki, S. and Deguchi, K. and Kawahara, G.},
  journal={J. Fluid Mech.},
  volume={1028},
  pages={A5},
  year={2026}
}

@book{Schlichting2016,
  title={{Boundary-Layer Theory}},
  author={Schlichting, H. and Gersten, K.},
  year={2016},
  publisher={Springer}
}

@book{Trefethen2000,
author = {Trefethen, L. N.},
title = {{Spectral Methods in MATLAB}},
publisher = {SIAM},
year = {2000},
}

@book{Boyd2000,
    AUTHOR = {Boyd, J. P.},
     TITLE = {{Chebyshev and Fourier Spectral Methods}},
 PUBLISHER = {Dover},
      YEAR = {2000},
   EDITION = {2nd}
}

@article{Hewitt2012,
  title={{Ultimate regime of high Rayleigh number convection in a porous medium}},
  author={Hewitt, D. R. and Neufeld, J. A. and Lister, J. R.},
  journal={Phys. Rev. Lett.},
  volume={108},
  pages={224503},
  year={2012},
}

@article{Blennerhassett1994,
  title={{Nonlinear high-wavenumber B{\'e}nard convection}},
  author={Blennerhassett, P. J. and Bassom, A. P.},
  journal={IMA J. Appl. Math.},
  volume={52},
  pages={51--77},
  year={1994}
}

@article{Hassanzadeh2014,
  title={Wall to wall optimal transport},
  author={Hassanzadeh, P. and Chini, G. P. and Doering, C. R.},
  journal={J. Fluid Mech.},
  volume={751},
  pages={627--662},
  year={2014},
}

@article{Souza2020,
  title={Wall-to-wall optimal transport in two dimensions},
  author={Souza, A. N. and Tobasco, I. and Doering, C. R.},
  journal={J. Fluid Mech.},
  volume={889},
  pages={A34},
  year={2020},
}

@article{Chilla2012,
author = {Chill{\`a}, F. and Schumacher, J.},
journal = {Eur. Phys. J. E},
title = {{New perspectives in turbulent Rayleigh--B{\'{e}}nard convection}},
volume = {35},
pages = {58},
year = {2012}
}

@article{Ahlers2009,
author = {Ahlers, G. and Grossmann, S. and Lohse, D.},
journal = {Rev. Mod. Phys.},
pages = {503--537},
title = {{Heat transfer and large scale dynamics in turbulent Rayleigh--B{\'{e}}nard convection}},
volume = {81},
year = {2009}
}

@book{Chandrasekhar1981,
author = {Chandrasekhar, S.},
publisher = {Dover},
title = {{Hydrodynamic and Hydromagnetic Stability}},
year = {1981}
}

\end{document}